\documentclass[pra,twocolumn,floatfix,a4paper,superscriptaddress]{revtex4}
\usepackage{bm,color,graphicx,amsmath,txfonts}
\usepackage{here}
\usepackage{array}
\usepackage{graphicx}
\usepackage[colorlinks=true,
citecolor=blue,
linkcolor=blue,
urlcolor=blue]{hyperref}
\usepackage{braket}
\usepackage{dsfont}
\usepackage[left=	2cm,top=2cm,right=1.2cm,bottom=2cm]{geometry}

\usepackage{tikz}
\everymath{\displaystyle}

\makeatletter
\renewcommand{\fnum@figure}{\textbf{Fig.~\thefigure:}}
\makeatother

\makeatletter
\renewcommand{\@makecaption}[2]{%
	\textbf{#1} #2\par
}
\makeatother
\makeatletter
\renewcommand{\fnum@figure}{\normalsize\textbf{Fig.~\thefigure:}}
\makeatother
\makeatletter
\renewcommand{\@makecaption}[2]{%
	\begin{flushleft}
		\textbf{#1} #2
	\end{flushleft}
}
\makeatother

\begin{document}
\makeatletter
\renewcommand{\@biblabel}[1]{%
	\makebox[2.1em][l]{\fontsize{10}{13}\selectfont[#1]}}
\makeatother

\title{Entanglement and Quantum Coherence in Coupled Double Quantum Dots under Markovian and Non-Markovian Noisy Channels}

\author{Omar Bachain}
\address{LPHE-Modeling and Simulation, Faculty of Sciences, Mohammed V University in Rabat, Rabat, Morocco}

\author{Mohamed \surname{Amazioug} }
\email{m.amazioug@uiz.ac.ma}
\address{LPTHE-Department of Physics, Faculty of Sciences, Ibnou Zohr University, Agadir 80000, Morocco}

\author{Nawal K. Almaymoni}
\address{Department of Physics, College of Science, Princess Nourah bint Abdulrahman University, P.O. Box 84428, Riyadh, 11671, Saudi Arabia}

\author{Rachid Ahl Laamara}
\address{LPHE-Modeling and Simulation, Faculty of Sciences, Mohammed V University in Rabat, Rabat, Morocco}
\address{Centre of Physics and Mathematics, CPM, Faculty of Sciences, Mohammed V University in Rabat, Rabat, Morocco}

\author{Naif S. Alharthi}
\address{Department of Physics, College of Sciences, University of Bisha, Bisha 61922, Saudi Arabia}

\author{Abdel-Haleem Abdel-Aty}
\address{Department of Physics, College of Sciences, University of Bisha, Bisha 61922, Saudi Arabia}

\date{\today}
\begin{abstract}
	
	Quantum dots are nanometer-scale semiconductor particles that exhibit size-dependent quantum mechanical properties.
	In this work, we investigate the dynamics of quantum correlations, quantified by the concurrence and the quantum coherence, in a bipartite system of coupled double quantum dots. The analysis is carried out within both Markovian and non-Markovian regimes, and further extended to different noisy quantum channels, including amplitude damping, phase flip, and phase damping.
	Our results show that environmental memory plays a crucial role in the preservation of quantum correlations, leading to oscillatory behavior and partial revivals in the non-Markovian regime, in contrast to the monotonic decay observed under Markovian dynamics. Moreover, distinct decoherence mechanisms induce qualitatively different effects: dissipative channels rapidly suppress correlations, while phase-based channels lead to either redistribution or gradual degradation.
	A key finding is that quantum coherence exhibits a higher robustness compared to entanglement under all considered conditions, highlighting its relevance as a reliable quantum resource in noisy environments. These results provide valuable insights into the control and protection of quantum correlations in realistic solid-state systems.
\end{abstract}
\maketitle
\section{Introduction}

Quantum correlations constitute one of the most fundamental and intriguing features of quantum mechanics, playing a central role in quantum information processing, quantum computation, and quantum communication \cite{Einstein1935,Bell1964,Bennett1993,Amico2008,Schrodinger1935,Benedetti2013,Amazioug2023,Jaloum2026,JaloumNPB2026}. Among these correlations, quantum entanglement has been widely recognized as a key resource enabling tasks that have no classical counterpart \cite{Wiseman2007}. However, it is now well established that entanglement does not exhaust all forms of quantum correlations, and more general quantities such as quantum coherence and quantum discord have attracted increasing attention in recent years.

\begin{figure}
	\centering
	\includegraphics[
	width=0.36\textwidth,
	trim=0 180 0 100,
	clip
	]{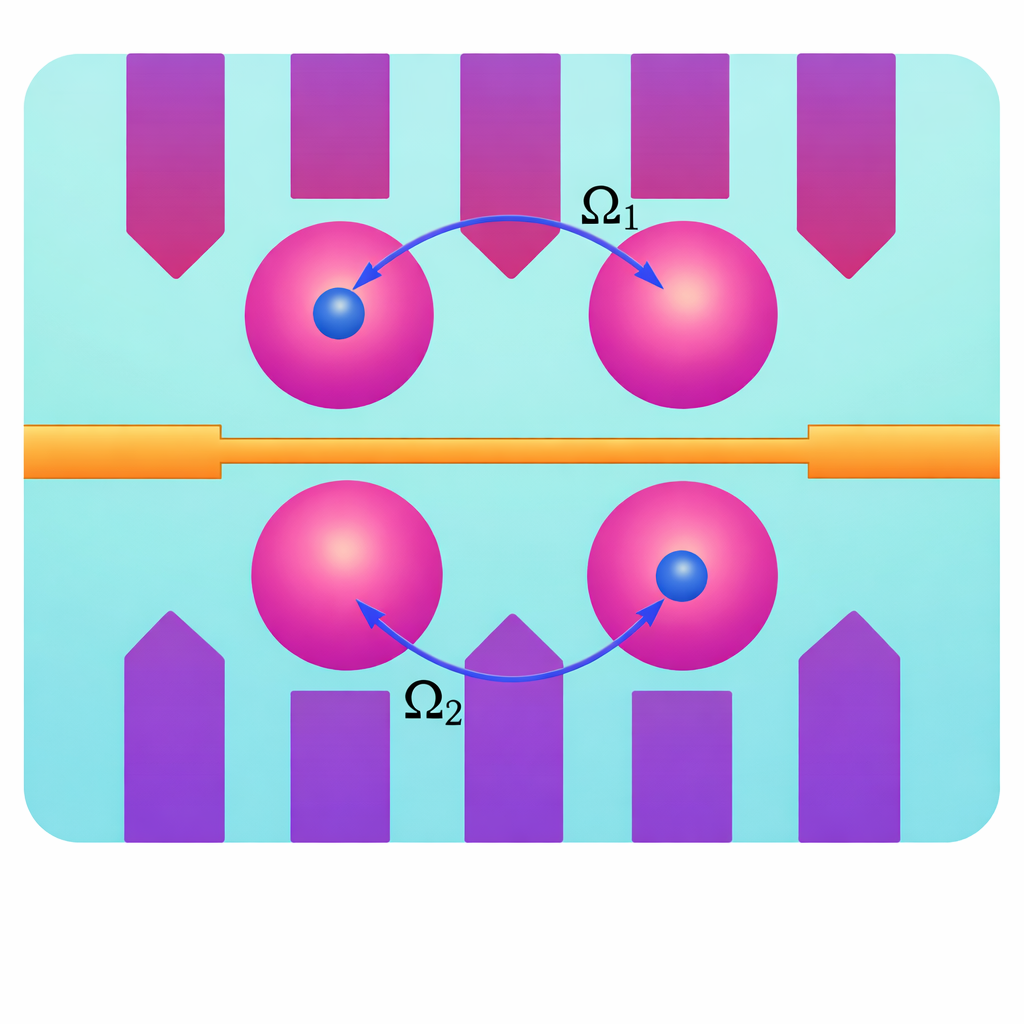}
	\caption{
		Schematic representation of the physical system consisting of two coupled double quantum dots. The electrons are depicted by the small spheres confined within each quantum dot. The tunneling couplings between the dots are characterized by $\Omega_1$ and $\Omega_2$.
	}
\end{figure}

In this context, quantum coherence \cite{Baumgratz2014}, arising from the superposition principle, has emerged as a fundamental resource in a broad range of physical applications, including quantum metrology, thermodynamics, and information processing \cite{Streltsov2017}. In particular, the notion of correlated coherence provides a unified framework to capture the interplay between coherence and quantum correlations, revealing that coherence can be redistributed between local and non-local contributions \cite{Tan2016}. Interestingly, it has been shown that correlated coherence can fully capture entanglement in certain regimes, especially at low temperatures, highlighting a deep connection between these two quantities.

On the other hand, semiconductor nanostructures, such as quantum dots and double quantum dots (DQDs), represent promising platforms for the physical implementation of quantum information devices due to their scalability and compatibility with existing electronic technologies \cite{Loss1998,Economou2012,Press2008,Gorman2005,Benito2017,DAnjou2019,Shi2012,Yang2020}. In particular, coupled double quantum dots can be modeled as charge qubits, where the position of an excess electron defines the logical states \cite{Shinkai2009,Fujisawa2004,Austing1998,Filgueiras2020}. These systems provide a versatile framework to investigate quantum correlations, as their properties can be tuned via tunneling amplitudes, Coulomb interactions, and external parameters such as temperature \cite{Filgueiras202}.

Despite their advantages, realistic quantum systems are inevitably affected by environmental interactions, leading to decoherence and the degradation of quantum correlations. This phenomenon has been extensively studied within the framework of open quantum systems \cite{BreuerPetruccione}, as well as in quantum dot platforms where mechanisms such as hyperfine interaction and charge noise play a crucial role \cite{Khaetskii2002,Coish2004,Petta2005}. Noisy quantum channels, including amplitude damping, phase flip, and phase damping \cite{Yu2004}, provide an effective theoretical description of these processes and enable a systematic investigation of the robustness of quantum resources under different physical conditions \cite{Palma1996}.

Motivated by these considerations, in this work we investigate the dynamics of quantum correlations, namely concurrence and quantum coherence, in a system of coupled double quantum dots subjected to various noisy channels. We analyze the influence of key physical parameters, including temperature, Coulomb interaction, and tunneling coupling, under both Markovian and non-Markovian regimes \cite{Breuer2009,Rivas2010,BreuerReview2016}. Particular attention is devoted to the comparison between entanglement and coherence, in order to identify their relative robustness and the conditions under which coherence can serve as an alternative quantum resource \cite{Bachain2026}.

Our results reveal that quantum coherence exhibits a higher robustness against environmental effects compared to entanglement, in agreement with previous studies \cite{Eleuch2017,Mohamed2022}, and provide further insight into the role of decoherence mechanisms in solid-state quantum systems.

The paper is organized as follows. In Sec.~\ref{sec2}, we introduce the physical model and the thermal density operator. Sec.~\ref{sec3} presents the measures of quantum correlations, namely concurrence and quantum coherence. In Sec.~\ref{sec4}, we investigate the dynamics in both Markovian and non-Markovian regimes. Sec.~\ref{sec5} is devoted to the analysis of noisy quantum channels, including amplitude damping, phase flip, and phase damping. The main results and discussion are presented in Sec.~\ref{sec6}. Sec.~\ref{sec7} discusses the experimental feasibility of the proposed model. Finally, the conclusions are given in Sec.~\ref{sec8}.

\section{Physical Model and Thermal Density Operator}\label{sec2}

We consider a bipartite quantum system composed of two coupled double quantum dots (DQDs), where each DQD contains a single excess electron and behaves as a charge qubit. The computational basis is defined as $\{|00\rangle, |01\rangle, |10\rangle, |11\rangle\}$, where $|0\rangle \equiv |L\rangle$ and $|1\rangle \equiv |R\rangle$ denote the electron localized in the left or right quantum dot, respectively. The Hamiltonian of the system is given by \cite{Fanchini2010,Filgueiras202}
\begin{equation}
\mathcal{H} = \Omega_1\, \tau_x^{(1)} + \Omega_2\, \tau_x^{(2)} + \mathcal{V} \left( \tau_z^{(1)} \otimes \tau_z^{(2)} \right),
\end{equation}
where $\tau_x$ and $\tau_z$ are Pauli matrices, $\Omega_1$ and $\Omega_2$ are the tunneling amplitudes, and $\mathcal{V}$ represents the Coulomb interaction between the electrons. The tunneling terms induce coherent superpositions between the localized states, while the interaction term introduces correlations between the qubits, leading to entanglement.

In the basis $\{|00\rangle, |01\rangle, |10\rangle, |11\rangle\}$, the Hamiltonian can be diagonalized analytically. By defining $N_{\pm} = \Omega_1 \pm \Omega_2$, the eigenvalues are given by
\begin{equation}
	{E}_{1,2} = \pm \sqrt{N_-^2 + \mathcal{V}^2}, \qquad
	E_{3,4} = \pm \sqrt{N_+^2 + \mathcal{V}^2}.
\end{equation}
The corresponding normalized eigenvectors are expressed as
\begin{align}\nonumber
	|\phi_1\rangle &= \Delta_- \left[ B_-(-|00\rangle + |11\rangle) + N_- (|01\rangle - |10\rangle) \right], \\\nonumber
	|\phi_2\rangle &= \Delta_- \left[ N_-(-|00\rangle + |11\rangle) + B_-(-|01\rangle + |10\rangle) \right], \\\nonumber
	|\phi_3\rangle &= \Delta_+ \left[ B_+ (|00\rangle + |11\rangle) + N_+ (|01\rangle + |10\rangle) \right], \\
	|\phi_4\rangle &= \Delta_+ \left[ N_+ (|00\rangle + |11\rangle) - B_+ (|01\rangle + |10\rangle) \right],
\end{align}
where
\begin{equation}
	\Delta_{\pm} = \frac{1}{\sqrt{2\left(N_{\pm}^2 + B_{\pm}^2\right)}}, \qquad
	B_{\pm} = \mathcal{V} + \sqrt{N_{\pm}^2 + \mathcal{V}^2}.
\end{equation}
These eigenvectors clearly show that the system eigenstates are entangled superpositions of charge configurations, resulting from the interplay between tunneling and Coulomb interaction.
At thermal equilibrium, the system is described by the Gibbs density operator
\begin{equation}
	\eta(T) = \frac{e^{-\beta \mathcal{H}}}{\mathcal{Z}},
\end{equation}
where $\beta = \frac{1}{k_B T}$ and the partition function is defined as
\begin{equation}
	\mathcal{Z} = \mathrm{Tr}\left[e^{-\beta \mathcal{H}}\right] = \sum_{i=1}^{4} e^{-\beta E_i}.
\end{equation}
By expanding $\eta(T)$ in the eigenbasis of the Hamiltonian, one obtains
\begin{equation}
	\eta(T) = \frac{1}{\mathcal{Z}} \sum_{i=1}^{4} e^{-\beta E_i} |\phi_i\rangle \langle \phi_i|.
\end{equation}
This expression shows that the thermal state is a statistical mixture of the eigenstates, where each state is weighted by its Boltzmann factor.
In the computational basis $\{|00\rangle, |01\rangle, |10\rangle, |11\rangle\}$, the density matrix takes the form
\begin{equation}
	\eta(T) =
	\begin{pmatrix}
		\eta_{11} & \eta_{12} & \eta_{13} & \eta_{14} \\
		\eta_{12} & \eta_{22} & \eta_{23} & \eta_{13} \\
		\eta_{13} & \eta_{23} & \eta_{22} & \eta_{12} \\
		\eta_{14} & \eta_{13} & \eta_{12} & \eta_{11}
	\end{pmatrix},\label{des matrix}
\end{equation}
where the matrix elements are given by
\begin{align}\nonumber
	\eta_{11} &= \frac{\Delta_-^2 \left(B_-^2 e^{-\beta E_1} + N_-^2 e^{-\beta E_2}\right) + \Delta_+^2 \left(B_+^2 e^{-\beta E_3} + N_+^2 e^{-\beta E_4}\right)}{\mathcal{Z}}, \\\nonumber
	\eta_{22} &= \frac{\Delta_-^2 \left(N_-^2 e^{-\beta E_1} + B_-^2 e^{-\beta E_2}\right) + \Delta_+^2 \left(N_+^2 e^{-\beta E_3} + B_+^2 e^{-\beta E_4}\right)}{\mathcal{Z}}, \\\nonumber
	\eta_{12} &= \frac{B_- N_- \Delta_-^2 \left(-e^{-\beta E_1} + e^{-\beta E_2}\right) + B_+ N_+ \Delta_+^2 \left(e^{-\beta E_3} - e^{-\beta E_4}\right)}{\mathcal{Z}}, \\\nonumber
	\eta_{13} &= \frac{B_- N_- \Delta_-^2 \left(e^{-\beta E_1} - e^{-\beta E_2}\right) + B_+ N_+ \Delta_+^2 \left(e^{-\beta E_3} - e^{-\beta E_4}\right)}{\mathcal{Z}}, \\\nonumber
	\eta_{14} &= \frac{-\Delta_-^2 \left(B_-^2 e^{-\beta E_1} + N_-^2 e^{-\beta E_2}\right) + \Delta_+^2 \left(B_+^2 e^{-\beta E_3} + N_+^2 e^{-\beta E_4}\right)}{\mathcal{Z}}, \\
	\eta_{23} &= \frac{-\Delta_-^2 \left(N_-^2 e^{-\beta E_1} + B_-^2 e^{-\beta E_2}\right) + \Delta_+^2 \left(N_+^2 e^{-\beta E_3} + B_+^2 e^{-\beta E_4}\right)}{\mathcal{Z}}.
\end{align}
The diagonal elements represent the populations of the computational states, while the off-diagonal elements encode the quantum coherence responsible for entanglement. At zero temperature, the system reduces to its ground state, whereas at finite temperature it becomes a mixed state due to thermal excitations. In the high-temperature limit, thermal fluctuations dominate and destroy quantum correlations. The Coulomb interaction $\mathcal{V}$ plays a crucial role, acting as a control parameter that enhances or suppresses entanglement depending on its strength.

\section{Quantum resources measures}\label{sec3}
\subsection{Quantum Entanglement }

To characterize the thermal entanglement of the system, we employ the concurrence, denoted by $\mathcal{C}$, which is a widely used measure for bipartite quantum correlations. \cite{Wootters1998,Hill1997}. This quantity provides a complete characterization of bipartite entanglement for two-qubit systems. It is defined as
\begin{equation}
	\mathcal{C} = \max \left\{ 0,\; \left| \sqrt{\upsilon_1} - \sqrt{\upsilon_3} \right| - \sqrt{\upsilon_2} - \sqrt{\upsilon_4} \right\},\label{conc}
\end{equation}
where $\upsilon_i \; (i = 1,2,3,4)$ are the eigenvalues, arranged in decreasing order, of the non-Hermitian matrix
\begin{equation}
	R = {\eta(T)} \, (\tau_y \otimes \tau_y)\, {\eta(T)}^{*} \, (\tau_y \otimes \tau_y),
\end{equation}
with $\tau_y$ denoting the Pauli matrix and ${\eta(T)}^{*}$ the complex conjugate of the density operator ${\eta(T)}$ in the computational basis $\{ |00\rangle, |01\rangle, |10\rangle, |11\rangle \}$.
In the context of the two coupled double quantum dots system under consideration, the evaluation of thermal entanglement requires the explicit determination of the eigenvalues of the matrix $R$. These eigenvalues can be expressed analytically as
\begin{align}\nonumber
	\upsilon_1 &= \Gamma_-^{\,2} + \frac{1}{2}\sqrt{\Xi_-^{\,2} - \Lambda_-^{\,2}}, \\\nonumber
	\upsilon_2 &= \Gamma_-^{\,2} - \frac{1}{2}\sqrt{\Xi_-^{\,2} - \Lambda_-^{\,2}}, \\\nonumber
	\upsilon_3 &= \Gamma_+^{\,2} + \frac{1}{2}\sqrt{\Xi_+^{\,2} - \Lambda_+^{\,2}}, \\
	\upsilon_4 &= \Gamma_+^{\,2} - \frac{1}{2}\sqrt{\Xi_+^{\,2} - \Lambda_+^{\,2}}.
\end{align}

The auxiliary quantities entering these expressions are defined in terms of the density matrix elements $\eta_{ij}$ as follows:
\begin{align}
	\Xi_{\pm} &= \left( \eta_{11} \pm \,\eta_{14} \right)^2 - \left( \eta_{22} \pm \,\eta_{23} \right)^2, \\
	\Lambda_{\pm} &= 2 \left( \eta_{13} \pm \eta_{14} \right)\left( \mp \eta_{11} - \,\eta_{14} + \,\eta_{23} \pm \eta_{22} \right), \\
\Gamma_{\pm} &= \left( \eta_{11} \pm \,\eta_{14} \right)^2 - 2^2\left( \eta_{12} \pm \eta_{13} \right)^2 + \left( \eta_{22} \pm \,\eta_{23} \right)^2.
\end{align}

It is important to emphasize that, due to the complexity of the density matrix elements for the considered system, the resulting analytical expression of the thermal concurrence becomes exceedingly cumbersome. Consequently, it is more appropriate to evaluate $\mathcal{C}$ numerically rather than presenting an explicit closed-form expression.

\subsection{Quantum Coherence}

Quantum coherence, regarded as a fundamental resource in quantum information processing, can be quantitatively characterized by means of the $l_1$-norm of coherence \cite{Baumgratz2014}. This measure captures the degree of quantum superposition by quantifying the distance between a given quantum state and the set of incoherent states.
For a bipartite quantum state ${\eta}(T) = \sum_{i,j} \eta_{ij}(T) |i\rangle \langle j|$, the $l_1$-norm of coherence is defined as the sum of the absolute values of all off-diagonal elements, namely
\begin{equation}
	\mathcal{C}_{l_1}({\eta}(T)) = \sum_{i \neq j} \left| \eta_{ij}(T) \right|.
\end{equation}

Accordingly, the quantum coherence of the considered two-qubit system is obtained directly from the off-diagonal elements of the density matrix as
\begin{equation}
	\mathcal{C}_{l_1}({\eta}(T)) = 2\,\Bigg[ 2\,\bigg( |\,\eta_{12}| + |\,\eta_{13}|\bigg)  +2\,\bigg(|\,\eta_{23}| +|\,\eta_{14}|\bigg)\Bigg] .\label{coher}
\end{equation}
\section{Dynamics of Open Quantum Systems: Markovian and Non-Markovian Regimes}\label{sec4}

In this section, we present the fundamental framework required to describe correlated quantum channels acting on a bipartite quantum system. Let us consider a two-qubit system initially prepared in the state ${\eta}(0)$. The evolution of this state under the action of a quantum channel $\mathcal{E}$ acting identically on both qubits can be described within the Kraus operator formalism as follows \cite{Hu2019,Hu2020}

\begin{equation}
	{\eta}(t) = \sum_{i,j=0}^{3} \mathcal{K}_{i,j} \, {\eta}(0) \, \mathcal{K}_{i,j}^\dagger,\label{Kraus}
\end{equation}

where the Kraus operators are defined by

\begin{equation}
	\mathcal{K}_{i,j} = \sqrt{P_{i,j}} \, \tau_i \otimes \tau_j,
\end{equation}

with $\tau_0 = \mathbb{I}$ and $\tau_{1,2,3} = \tau_{x,y,z}$ denoting the identity and the Pauli operators, respectively. The coefficients $P_{i,j}$ define a joint probability distribution satisfying

\begin{equation}
	P_{i,j} \geq 0, \quad \sum_{i,j} P_{i,j} = 1,
\end{equation}

and the completeness relation ensuring complete positivity and trace preservation (CPTP condition) reads

\begin{equation}
	\sum_{i,j} \mathcal{K}_{i,j}^\dagger \mathcal{K}_{i,j} = \mathbb{I}.
\end{equation}

To incorporate classical correlations between successive applications of the channel, the joint probabilities are modeled as \cite{Macchiavello2002}

\begin{equation}
	P_{i,j} = (1-\mu)\, P_i P_j + \mu \, P_i \delta_{i,j},
\end{equation}

where $\delta_{i,j}$ is the Kronecker delta and $\mu \in [0,1]$ quantifies the degree of classical correlations. The limiting cases $\mu = 0$ and $\mu = 1$ correspond to uncorrelated and fully correlated channels, respectively.
In this work, we focus on a correlated dephasing channel characterized by the probability distribution

\begin{equation}
	P_0 = 1 - P, \quad P_1 = P_2 = 0, \quad P_3 = P.
\end{equation}
To describe the time-dependent decoherence process, we consider a colored pure dephasing model governed by the stochastic Hamiltonian \cite{Daffer2004}
\begin{equation}
	\hat{H}(t) = \Delta(t) \, \sigma_z,
\end{equation}
where $\Delta(t)$ is a stochastic variable modeled as a random telegraph signal,
\begin{equation}
	\Delta(t) = \alpha \, m(t),
\end{equation}
with $m(t)$ following a Poisson process such that $\langle m(t) \rangle = {1}/{2\tau}$, and $\alpha = \pm 1$ is a dichotomic random variable.
The time-dependent decoherence function $p(t)$ can then be expressed as
\begin{equation}
	P(t) = \frac{1 - F(t)}{2}.
\end{equation}

The function $F(t)$ distinguishes between Markovian and non-Markovian regimes

\paragraph{Non-Markovian regime ($4\tau > 1$):}

\begin{equation}
	F(t) = e^{-t/(2\tau)} \left[ \cos\left(\frac{v t}{2\tau}\right) + \frac{1}{v} \sin\left(\frac{v t}{2\tau}\right) \right],
\end{equation}

with
	$v = \sqrt{|1 - 4\tau^2|}$.
	\vspace{0.2cm}
\paragraph{Markovian regime ($4\tau < 1$):}
\begin{equation}
	F(t) = e^{-t/(2\tau)} \left[ \cosh\left(\frac{v t}{2\tau}\right) + \frac{1}{v} \sinh\left(\frac{v t}{2\tau}\right) \right].
\end{equation}
We now consider the evolution of an initial thermal state ${\eta}(T)$ under the correlated dephasing channel. By applying the Kraus map, the time-dependent density operator takes the form

\begin{equation}
	\eta(t) =
	\begin{pmatrix}
		\eta_{11} & \gamma\,\eta_{12} & \gamma\,\eta_{13} & \gamma\,\eta_{14} \\
		\gamma\,\eta_{12} & \eta_{22} & \gamma\,\eta_{23} & \gamma\,\eta_{13} \\
	\gamma\,	\eta_{13} &\gamma\,\eta_{23} & \eta_{22} & \gamma\,\eta_{12} \\
		\gamma\,\eta_{14} &\gamma\,\eta_{13} &\gamma\, \eta_{12} & \eta_{11}
	\end{pmatrix},
\end{equation}
where the decoherence factor $\gamma$ is given by
\begin{equation}
	\gamma = (1-\mu)\, F^2(t) + \mu.
\end{equation}

This expression clearly shows that the off-diagonal elements of the density matrix are affected by the decoherence process, while the diagonal populations remain invariant. The parameter $\mu$ plays a crucial role in preserving quantum coherence: increasing classical correlations ($\mu \to 1$) leads to a suppression of decoherence effects, thereby enhancing the robustness of quantum correlations. As a direct consequence, both the concurrence and the quantum coherence acquire the same time dependence through the decoherence factor $\gamma(t)$, since the off-diagonal elements entering their definitions are simply rescaled as $\rho_{ij}(t)=\gamma\,\eta_{ij}(T)$ $(i\neq j)$. It is worth noting that the expressions of the concurrence and the quantum coherence are obtained by multiplying the off-diagonal elements of the density matrix by the decoherence factor $\gamma(t)$, while the diagonal elements remain unchanged.

\section{Decoherence Channels: Amplitude Damping, Phase Flip, and Phase Damping}\label{sec5}

The interaction of a quantum system with its surrounding environment leads to decoherence, resulting in the degradation of quantum coherence and quantum correlations. To model these effects, several paradigmatic noisy quantum channels are commonly employed, including the amplitude damping (AD), phase flip (PF), and phase damping (PD) channels \cite{Nielsen2010,Pirandola2008,Damodarakurup2009,AbdRabbou2022}.
The dynamical evolution of the system under these channels is described by the Kraus operator formalism introduced in Eq.~(\ref{Kraus}), where the two-qubit Kraus operators are given by $\mathcal{K}_{kl} = \mathcal{K}_k \otimes \mathcal{K}_l$.
In what follows, we analyze the effect of these channels on the initial state defined in the previous section.

\subsection{Amplitude Damping Channel}

The amplitude damping (AD) channel describes dissipative processes such as spontaneous emission. The single-qubit Kraus operators are given by

\begin{equation}
\mathcal{K}_1 =
	\begin{pmatrix}
		1 & 0 \\
		0 & \sqrt{1-s}
	\end{pmatrix}, \quad
	\mathcal{K}_2 =
	\begin{pmatrix}
		0 & \sqrt{s} \\
		0 & 0
	\end{pmatrix},\label{exp Kraus}
\end{equation}

The decoherence parameter is defined as $s = 1 - e^{-vt}$, with $s \in [0,1]$ and $v$ denoting the decay rate. By substituting Eqs.~(\ref{des matrix}) and (\ref{exp Kraus}) into Eq.~(\ref{Kraus}), one obtains the explicit form of the density matrix describing the system under the amplitude damping (AD) channel:

\begin{equation}
	{\eta}^{AD} =
	\begin{pmatrix}
		\delta_{11} & \delta_{12} & \delta_{13} & \delta_{14} \\
		\delta_{12} & \delta_{22} & \delta_{23} & \delta_{13} \\
		\delta_{13} & \delta_{23} & \delta_{22} & \delta_{12} \\
		\delta_{14} & \delta_{13} & \delta_{12} & \delta_{11}
	\end{pmatrix},
\end{equation}
with
\begin{align}\nonumber
	\delta_{11} &= \eta_{11} + s^2\, \eta_{11} + 2 s \,\eta_{22},\quad
	\delta_{22} =  -(1-s)(s\,\rho_{11}+\eta_{22}), \\
	\delta_{12} &= \sqrt{1-s}(1+s)\,\eta_{12},\qquad\nonumber
	\delta_{13} = \sqrt{1-s}(1+s)\,\eta_{13}, \\\nonumber
	\delta_{14} &= (1-s)\,\rho_{14},\qquad\qquad\,\,\,\,\,
	\delta_{23} = (1-s)\,\eta_{23}.
\end{align}
The expressions of the concurrence and the coherence can be directly obtained by substituting $\eta_{ij}$ with $\delta_{ij}$ in Eqs.~(\ref{conc}) and (\ref{coher}), respectively.
\subsection{Phase Flip Channel}

The phase flip (PF) channel describes phase errors induced by environmental fluctuations. The corresponding Kraus operators are
\begin{equation}
	\mathcal{K}_1 =
	\begin{pmatrix}
		\sqrt{s} & 0 \\
		0 & \sqrt{s}
	\end{pmatrix}, \quad
	\mathcal{K}_2 =
	\begin{pmatrix}
		\sqrt{1-s} & 0 \\
		0 & -\sqrt{1-s}
	\end{pmatrix}.
\end{equation}

Under the action of this channel, the density matrix becomes

\begin{equation}
	{\eta}^{PF} =
	\begin{pmatrix}
		\pi_{11} & \pi_{12} & \pi_{13} & \pi_{14} \\
		\pi_{12} & \pi_{22} & \pi_{23} & \pi_{13} \\
		\pi_{13} & \pi_{23} & \pi_{22} & \pi_{12} \\
		\pi_{14} & \pi_{13} & \pi_{12} & \pi_{11}
	\end{pmatrix},
\end{equation}
where
\begin{align}\nonumber
	\pi_{11} &= \eta_{11},\qquad\qquad\quad\,\,\,\,\,\,
	\pi_{22} =  \eta_{22} \\\nonumber
	\pi_{12} &=(-1 + 2 s) \,\eta_{12},\qquad \nonumber
	\pi_{13} = (-1 + 2 s) \,\eta_{13}, \\\nonumber
	\pi_{14} &= (1 - 2 s)^2\,\rho_{14},\qquad\,\,
	\pi_{23} = (1 - 2 s)^2\,\eta_{23}.
\end{align}
The concurrence and coherence follow directly from Eqs.~(\ref{conc}) and (\ref{coher}) by replacing $\eta_{ij}$ with $\pi_{ij}$, respectively.
\subsection{Phase Damping Channel}

The phase damping (PD) channel describes pure dephasing processes without energy exchange. The corresponding Kraus operators are

\begin{equation}
	\mathcal{K}_1 =
	\begin{pmatrix}
		1 & 0 \\
		0 & \sqrt{1-s}
	\end{pmatrix}, \quad
	\mathcal{K}_2 =
	\begin{pmatrix}
		0 & 0 \\
		0 & \sqrt{s}
	\end{pmatrix}.
\end{equation}

The evolved density matrix is then given by

\begin{equation}
	{\eta}^{PD} =
	\begin{pmatrix}
		\varphi_{11} & \varphi_{12} & \varphi_{13} & \varphi_{14} \\
		\varphi_{12} & \varphi_{22} & \varphi_{23} & \varphi_{13} \\
		\varphi_{13} & \varphi_{23} & \varphi_{22} & \varphi_{12} \\
		\varphi_{14} & \varphi_{13} & \varphi_{12} & \varphi_{11}
	\end{pmatrix},
\end{equation}
with
\begin{align}\nonumber
	\varphi_{11} &= \eta_{11},\qquad\qquad\quad
	\varphi_{22} =  \eta_{22} \\\nonumber
	\varphi_{12} &=\sqrt{1 - s}\,\eta_{12},\qquad \nonumber
	\varphi_{13} = \sqrt{1 - s} \,\eta_{13}, \\\nonumber
	\varphi_{14} &= (1 -  s)\,\rho_{14},\qquad\,\,
	\varphi_{23} = (1 - s)\,\eta_{23}.
\end{align}

In a similar manner, the concurrence and coherence can be derived from Eqs.~(\ref{conc}) and (\ref{coher}) via the substitution $\eta_{ij} \to \pi_{ij}$, respectively.

\section{Results and Discussions}\label{sec6}

In this section, we explore the dynamical behavior of quantum correlations, focusing on entanglement and quantum coherence under correlated noisy channels. We begin with the Markovian regime and then extend our analysis to the non-Markovian case. For both regimes, we examine the influence of amplitude damping (AD), phase flip (PF), and phase damping (PD) channels. We further analyze the role of key physical parameters, including temperature, Coulomb interaction, and tunneling couplings. The obtained results are discussed in detail to reveal the distinct effects induced by each decoherence process.

\subsection{Markovian and non-Markovian regimes}

\begin{figure*}[t!]
	\includegraphics[width=0.33\linewidth]{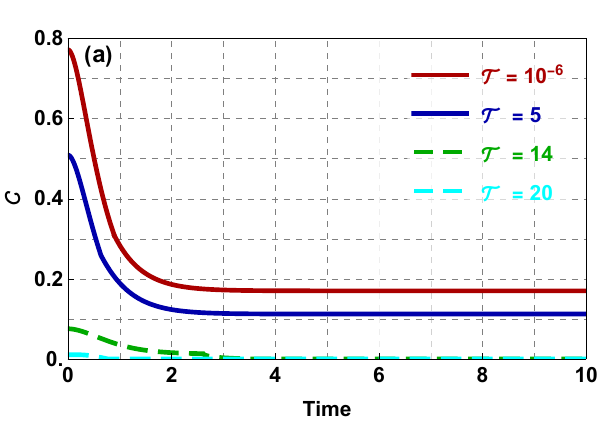}
	\includegraphics[width=0.33\linewidth]{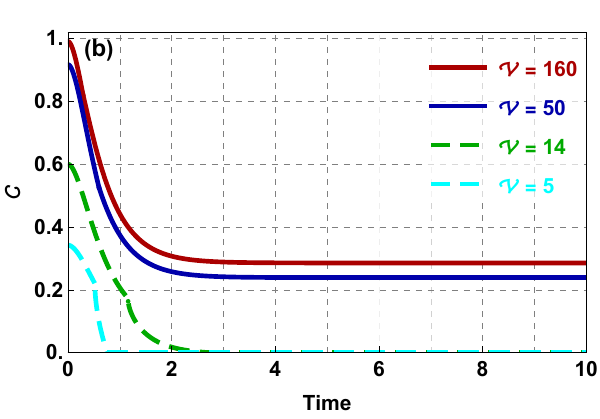}
	\includegraphics[width=0.33\linewidth]{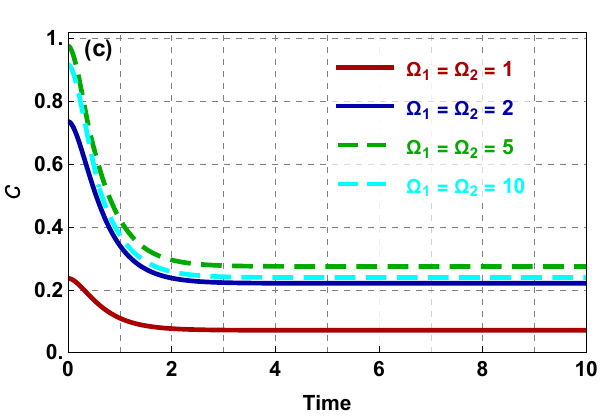}
\caption{
	Dynamical evolution of the concurrence $\mathcal{C}$ in the Markovian regime with $\tau = 0.2$ and $\mu = 0.3$. 
	(a) For different values of the temperature $\mathcal{T}$, with $\Omega_1 = 10$, $\Omega_2 = 15$, and $\mathcal{V} = 25$. 
	(b) For different values of the  Coulomb potential $\mathcal{V}$, with $\mathcal{T} = 0.1$, $\Omega_1 = 10$, and $\Omega_2 = 15$. 
	(c) For different values  of coupling $\Omega_1 = \Omega_2$, with $\mathcal{T} = 0.1$ and $\mathcal{V} = 40$.
}
	\label{fig1}
\end{figure*}
Figure~\ref{fig1} illustrates the time evolution of the concurrence $\mathcal{C}$ in the Markovian regime, highlighting the influence of temperature, Coulomb interaction, and tunneling coupling on the entanglement dynamics.
In Fig.~\ref{fig1}(a), the effect of temperature is clearly observed. For high temperatures ($\mathcal{T}=14$ and $\mathcal{T}=20$), the entanglement rapidly decays and vanishes within a short time, indicating the destructive role of thermal fluctuations on quantum correlations. In contrast, for low temperatures ($\mathcal{T}=10^{-6}$ and $\mathcal{T}=5$), the concurrence initially decreases but then reaches a non-zero steady value, remaining approximately constant over time. This behavior reflects the persistence of residual entanglement when thermal excitations are weak, allowing quantum correlations to survive despite the Markovian decoherence process.
Figure~\ref{fig1}(b) shows a similar qualitative behavior when varying the Coulomb interaction $\mathcal{V}$. For weak interaction strengths, the entanglement decays more rapidly and may disappear completely. However, as $\mathcal{V}$ increases, the concurrence becomes more robust and stabilizes at a finite value after an initial decay. This demonstrates that the Coulomb interaction plays a constructive role in protecting entanglement against environmental effects by strengthening the correlations between the qubits.
In Fig.~\ref{fig1}(c), the influence of the tunneling coupling $\Omega_1=\Omega_2$ is analyzed. The concurrence exhibits an initial decay followed by a saturation toward a constant value at long times. Increasing the coupling strength enhances the stationary value of entanglement, indicating that stronger tunneling promotes the formation and preservation of quantum correlations. This behavior can be attributed to the increased coherent superposition between quantum states induced by the tunneling terms in the Hamiltonian.
Overall, Fig.~\ref{fig1} demonstrates that while Markovian decoherence leads to a monotonic degradation of entanglement, its long-time behavior strongly depends on the system parameters. In particular, low temperature, strong Coulomb interaction, and large tunneling coupling all contribute to enhancing the robustness of quantum entanglement.
\begin{figure*}[t!]
	\includegraphics[width=0.33\linewidth]{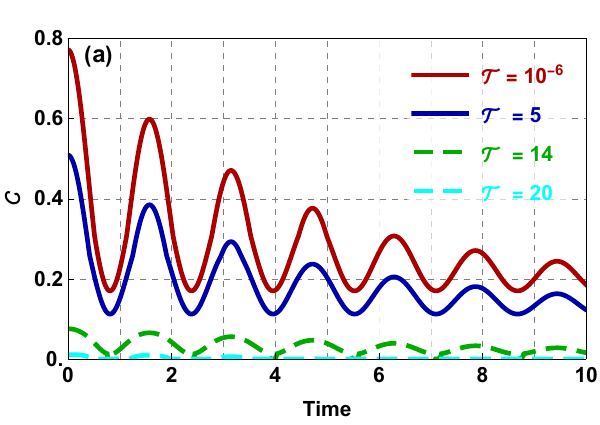}
	\includegraphics[width=0.33\linewidth]{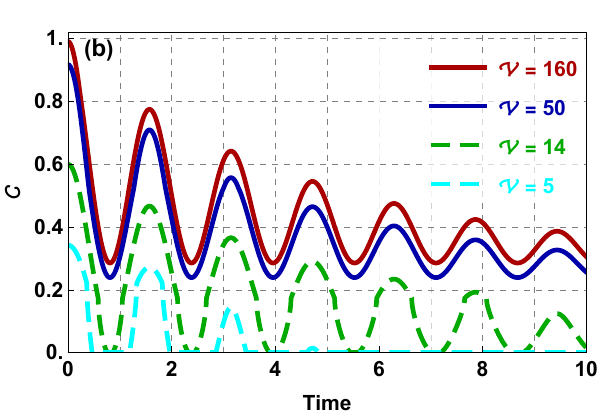}
	\includegraphics[width=0.33\linewidth]{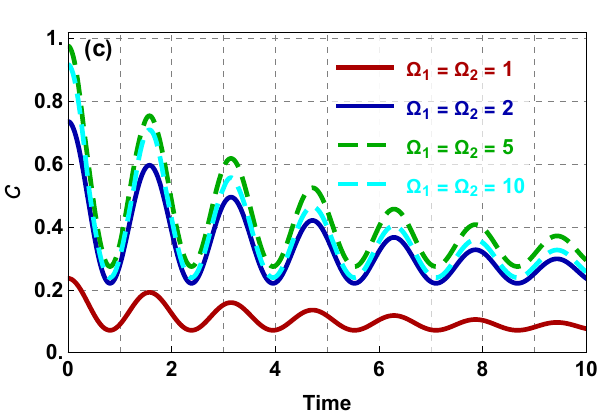}
	\caption{
		Dynamical evolution of the concurrence $\mathcal{C}$ in the Non-Markovian regime with $\tau = 5$ and $\mu = 0.3$. 
		(a) For different values of the temperature $\mathcal{T}$, with $\Omega_1 = 10$, $\Omega_2 = 15$, and $\mathcal{V} = 25$. 
		(b) For different values of the  Coulomb potential $\mathcal{V}$, with $\mathcal{T} = 0.1$, $\Omega_1 = 10$, and $\Omega_2 = 15$. 
		(c) For different values  of coupling $\Omega_1 = \Omega_2$, with $\mathcal{T} = 0.1$ and $\mathcal{V} = 40$.
	}
	\label{fig2}
\end{figure*}
Figure~\ref{fig2} presents the dynamics of the concurrence $\mathcal{C}$ in the non-Markovian regime, where memory effects of the environment play a crucial role in the evolution of quantum correlations.
In Fig.~\ref{fig2}(a), the dynamics is strongly influenced by temperature through a competition between thermal fluctuations and memory-induced information backflow. For high temperatures ($\mathcal{T}=14$ and $\mathcal{T}=20$), the entanglement is rapidly suppressed, and the revival effects remain weak, indicating that thermal noise dominates over non-Markovian memory. Conversely, at low temperatures ($\mathcal{T}=10^{-6}$ and $\mathcal{T}=5$), pronounced oscillations appear in the concurrence, revealing clear revivals of entanglement. This behavior is a direct signature of information flowing back from the environment to the system, partially restoring quantum correlations.
Figure~\ref{fig2}(b) highlights the role of the Coulomb interaction $\mathcal{V}$ in shaping the non-Markovian dynamics. For small values of $\mathcal{V}$, the revival amplitudes are limited, and the entanglement remains fragile. As $\mathcal{V}$ increases, the oscillatory behavior becomes more pronounced, with larger revival peaks and a higher average concurrence. This indicates that stronger inter-qubit interaction enhances the system’s ability to store and retrieve quantum information from the environment, thus amplifying non-Markovian effects.
In Fig.~\ref{fig2}(c), varying the tunneling coupling $\Omega_1=\Omega_2$ modifies the coherence exchange within the system. For weak coupling, the entanglement exhibits small-amplitude oscillations, reflecting limited coherence redistribution. Increasing the coupling strength leads to more regular and higher-amplitude oscillations, indicating a more efficient interplay between internal coherent dynamics and environmental memory. As a result, the system displays sustained entanglement revivals over time.
Overall, the non-Markovian regime is characterized by the emergence of entanglement revivals and oscillatory behavior, in contrast to the monotonic decay observed in the Markovian case. These results demonstrate that memory effects, combined with suitable physical parameters, can significantly enhance the resilience of quantum correlations.

\begin{figure*}[t!]
	\includegraphics[width=0.33\linewidth]{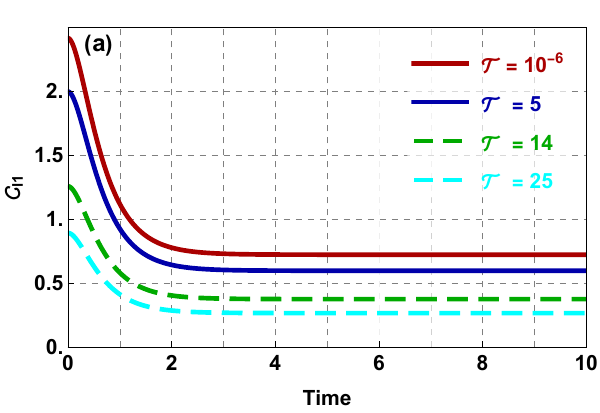}
	\includegraphics[width=0.33\linewidth]{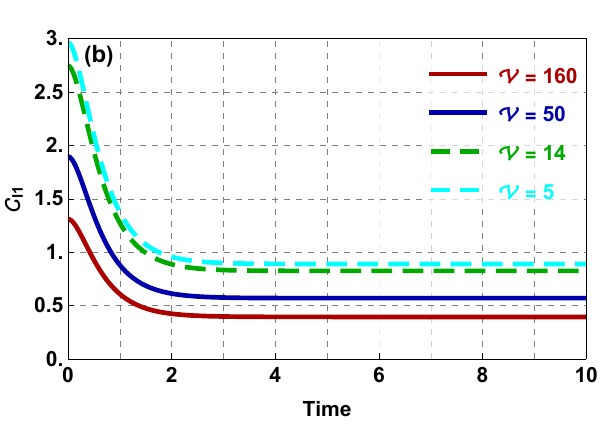}
	\includegraphics[width=0.33\linewidth]{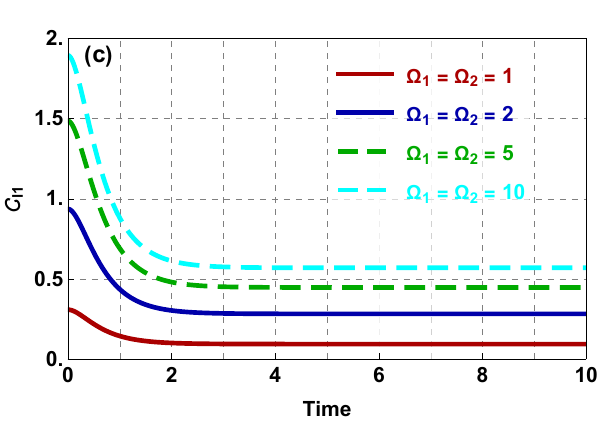}
	\caption{
		Dynamical evolution of the the quantum coherence $\mathcal{C}_{l_1}$ in the Markovian regime with $\tau = 0.2$ and $\mu = 0.3$. 
		(a) For different values of the temperature $\mathcal{T}$, with $\Omega_1 = 10$, $\Omega_2 = 15$, and $\mathcal{V} = 25$. 
		(b) For different values of the  Coulomb potential $\mathcal{V}$, with $\mathcal{T} = 0.1$, $\Omega_1 = 10$, and $\Omega_2 = 15$. 
		(c) For different values  of coupling $\Omega_1 = \Omega_2$, with $\mathcal{T} = 0.1$ and $\mathcal{V} = 40$.
	}
	\label{fig3}
\end{figure*}
Figure~\ref{fig3} depicts the time evolution of the quantum coherence $\mathcal{C}_{l_1}$ in the Markovian regime. Unlike entanglement, coherence quantifies the amount of quantum superposition present in the system and provides complementary insight into the degradation of quantum resources.
In Fig.~\ref{fig3}(a), coherence decreases progressively over time due to the irreversible loss of phase information induced by the Markovian environment. Although increasing temperature accelerates this decay, the coherence remains finite over a relatively long time scale, confirming its robustness compared to entanglement.
Figure~\ref{fig3}(b) shows that increasing the Coulomb interaction $\mathcal{V}$ leads to a reduction of quantum coherence. This behavior can be understood by noting that strong interaction tends to localize the system in more correlated configurations, thereby suppressing quantum superpositions between different basis states. As a result, the off-diagonal elements of the density matrix decrease, leading to a lower value of $\mathcal{C}_{l_1}$.
In Fig.~\ref{fig3}(c), the tunneling coupling $\Omega_1=\Omega_2$ enhances the creation of coherent superpositions. Larger coupling strengths lead to higher initial coherence, followed by a smooth decay toward a finite stationary value. This indicates that tunneling processes counteract decoherence by maintaining quantum superpositions.
Overall, Fig.~\ref{fig3} highlights that quantum coherence, although affected by environmental noise, remains more resilient than entanglement. Its evolution is governed by a balance between interaction-induced localization and tunneling-induced delocalization.

\begin{figure*}[t!]
	\includegraphics[width=0.33\linewidth]{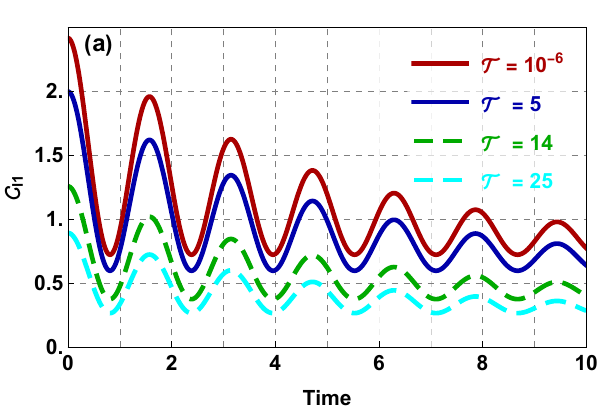}
	\includegraphics[width=0.33\linewidth]{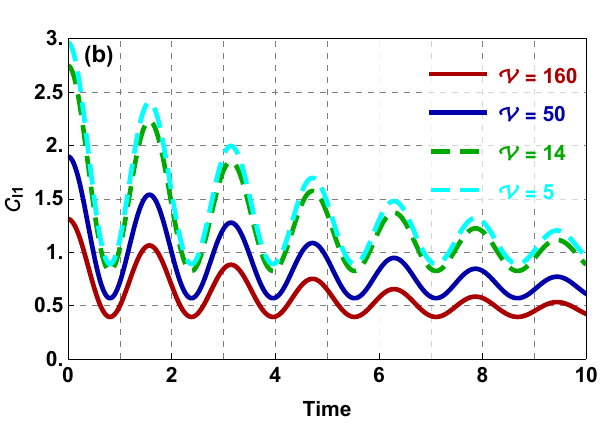}
	\includegraphics[width=0.33\linewidth]{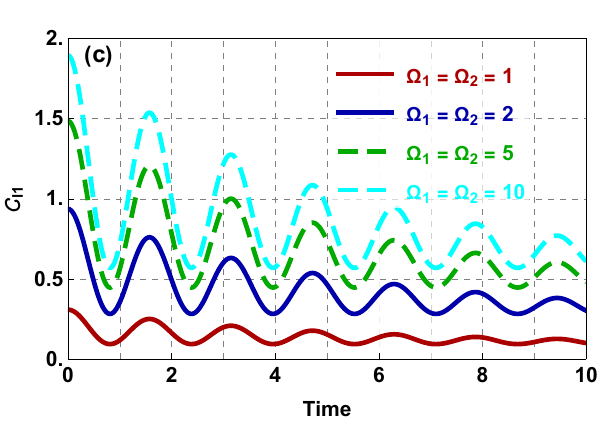}
\caption{
	Dynamical evolution of the the quantum coherence $\mathcal{C}_{l_1}$ in the Non-Markovian regime with $\tau = 5$ and $\mu = 0.3$. 
	(a) For different values of the temperature $\mathcal{T}$, with $\Omega_1 = 10$, $\Omega_2 = 15$, and $\mathcal{V} = 25$. 
	(b) For different values of the  Coulomb potential $\mathcal{V}$, with $\mathcal{T} = 0.1$, $\Omega_1 = 10$, and $\Omega_2 = 15$. 
	(c) For different values  of coupling $\Omega_1 = \Omega_2$, with $\mathcal{T} = 0.1$ and $\mathcal{V} = 40$.
}
	\label{fig4}
\end{figure*}

Figure~\ref{fig4} reveals a markedly different behavior of quantum coherence in the non-Markovian regime, where memory effects induce a non-trivial exchange of information between the system and its environment.
A first striking feature, visible in Fig.~\ref{fig4}(a), is the emergence of oscillations in $\mathcal{C}_{l_1}$, which reflect recurrent partial restoration of coherence. At low temperatures ($\mathcal{T}=10^{-6}$ and $\mathcal{T}=5$), these oscillations are clearly visible and persist over long times, indicating that the environment temporarily feeds coherence back into the system. As the temperature increases, this revival mechanism becomes progressively suppressed, and the dynamics tends toward a damped profile dominated by thermal noise.
The role of the Coulomb interaction $\mathcal{V}$, illustrated in Fig.~\ref{fig4}(b), appears through a modulation of both the amplitude and the persistence of these oscillations. Larger values of $\mathcal{V}$ tend to reduce the overall level of coherence, consistent with an interaction-induced localization that weakens superposition effects. At the same time, the oscillatory pattern remains present, showing that non-Markovian memory effects survive even when coherence is partially suppressed.
A different mechanism emerges when tuning the tunneling coupling $\Omega_1=\Omega_2$, as shown in Fig.~\ref{fig4}(c). Increasing the coupling enhances the exchange of coherence between the qubits, leading to more regular and pronounced oscillations. In this case, the internal coherent dynamics and the environmental memory act cooperatively, allowing the system to sustain a significant amount of coherence over time.
These results underline that, in contrast to the Markovian regime, coherence dynamics is no longer governed by a simple decay process but rather by a continuous competition between decoherence, interaction-induced localization, and memory-driven recoherence effects.
\begin{figure}
	\centering
	\includegraphics[width=0.92\linewidth]{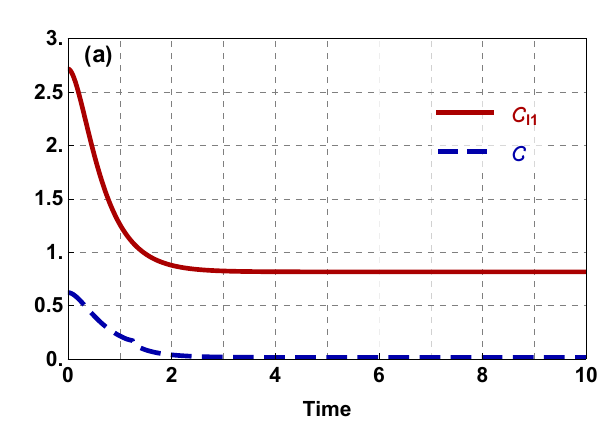}\\
	\includegraphics[width=0.92\linewidth]{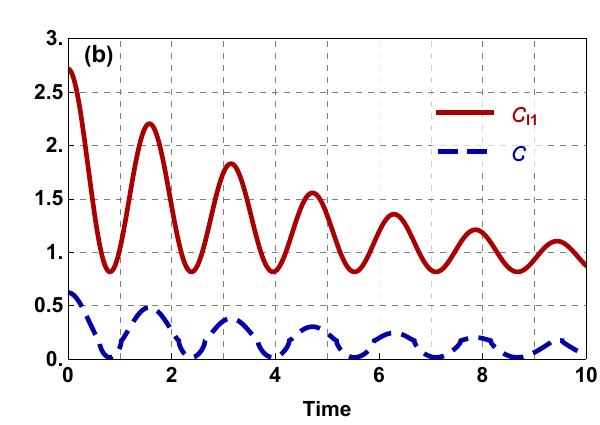}
	\caption{
		Comparison between the concurrence $\mathcal{C}$ and the quantum coherence $\mathcal{C}_{l_1}$. 
		(a) Markovian regime with $\tau = 0.2$ and $\mu = 0.3$. 
		(b) Non-Markovian regime with $\tau = 0.2$ and $\mu = 0.3$. 
		The remaining parameters are fixed at $\mathcal{T} = 0.01$, $\mathcal{V} = 15$, $\Omega_1 = 10$, and $\Omega_2 = 15$.
	}
	\label{fig5}
\end{figure}

Figure~\ref{fig5} provides a direct comparison between the concurrence $\mathcal{C}$ and the quantum coherence $\mathcal{C}_{l_1}$, offering deeper insight into the distinct behavior of these two quantum resources under decoherence.
As observed in Fig.~\ref{fig5}(a), corresponding to the Markovian regime, both quantities exhibit a monotonic decay due to the irreversible loss of quantum information. However, a clear distinction emerges: the concurrence decreases more rapidly and may vanish completely after a finite time, whereas the coherence $\mathcal{C}_{l_1}$ decays more gradually and remains non-zero over a longer time scale. This indicates that coherence is more robust than entanglement against Markovian noise, as it only requires the presence of quantum superpositions, while entanglement relies on stronger non-classical correlations 
A qualitatively different scenario is revealed in Fig.~\ref{fig5}(b), where non-Markovian effects come into play. In this regime, both concurrence and coherence exhibit oscillatory behavior associated with information backflow from the environment. Nevertheless, coherence still dominates in magnitude and persists over longer times, while entanglement undergoes more pronounced collapses and revivals. This highlights the greater sensitivity of entanglement to environmental disturbances, even in the presence of memory effects :contentReference[oaicite:1]{index=1}.
This comparison clearly demonstrates that quantum coherence constitutes a more resilient resource than entanglement under both Markovian and non-Markovian dynamics. While entanglement is more fragile and can suffer sudden death, coherence survives longer and exhibits smoother dynamics, making it a valuable resource for quantum information processing in noisy environments.

In summary, the dynamics of entanglement and coherence strongly depends on the nature of the environment. In the Markovian regime, both quantities exhibit a monotonic decay due to the irreversible loss of information. In contrast, the non-Markovian regime is characterized by oscillations and partial revivals induced by memory effects.
Notably, quantum coherence remains more robust than entanglement, persisting over longer times. These observations motivate the analysis of their behavior under different noisy quantum channels.

\subsection{Effects of Noisy Quantum Channels}
\begin{figure*}[t!]
	\includegraphics[width=0.33\linewidth]{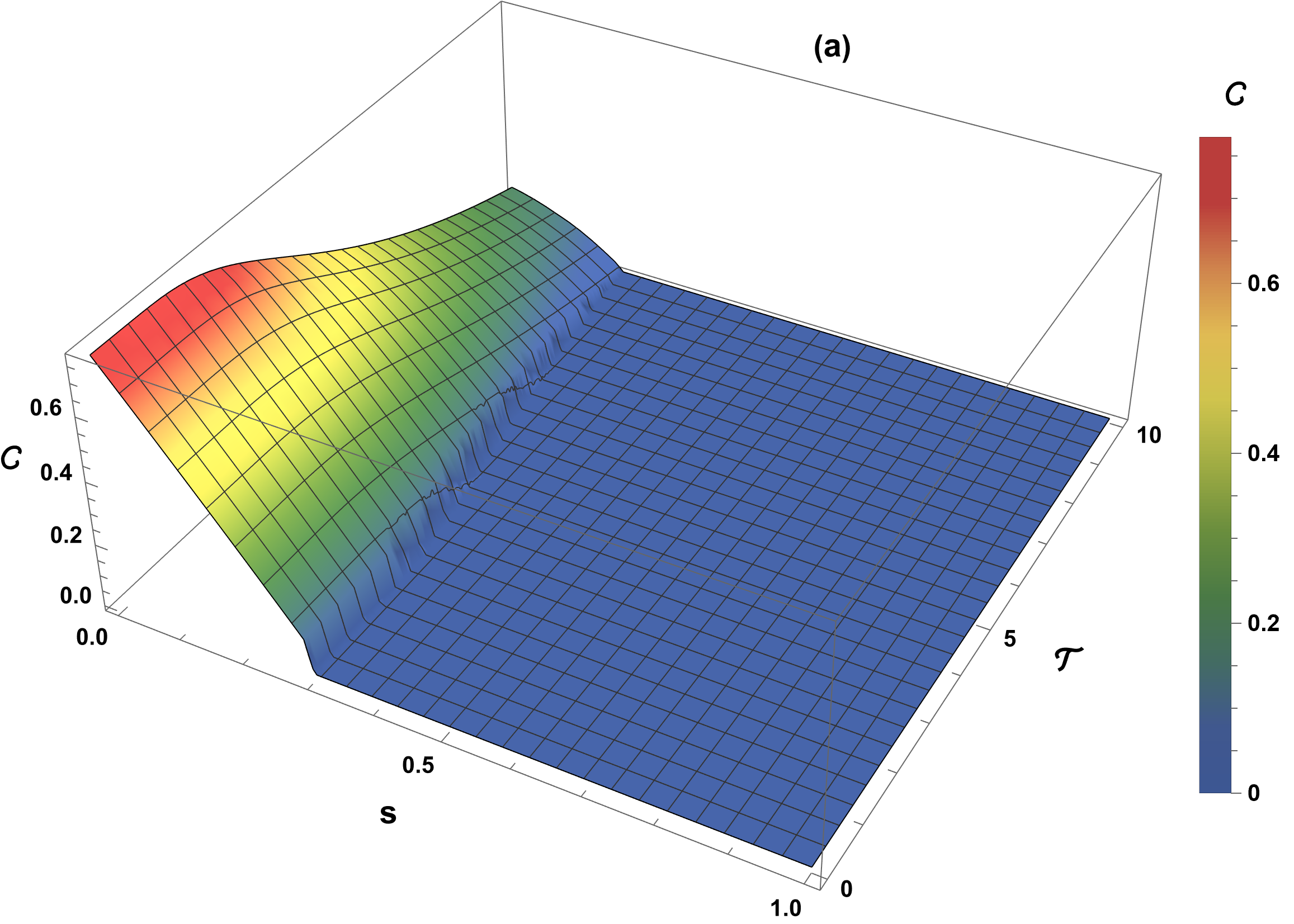}
	\includegraphics[width=0.33\linewidth]{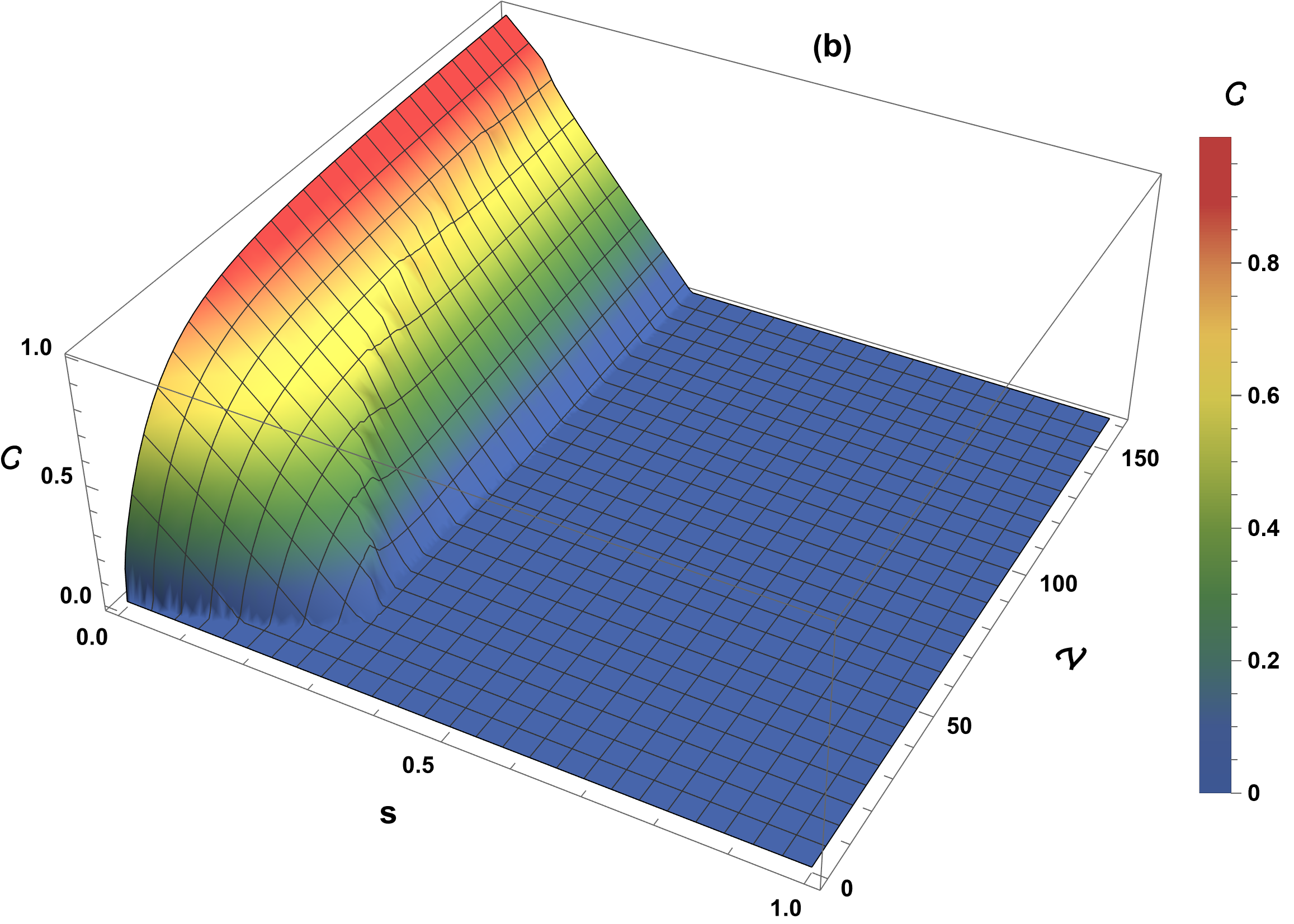}
	\includegraphics[width=0.33\linewidth]{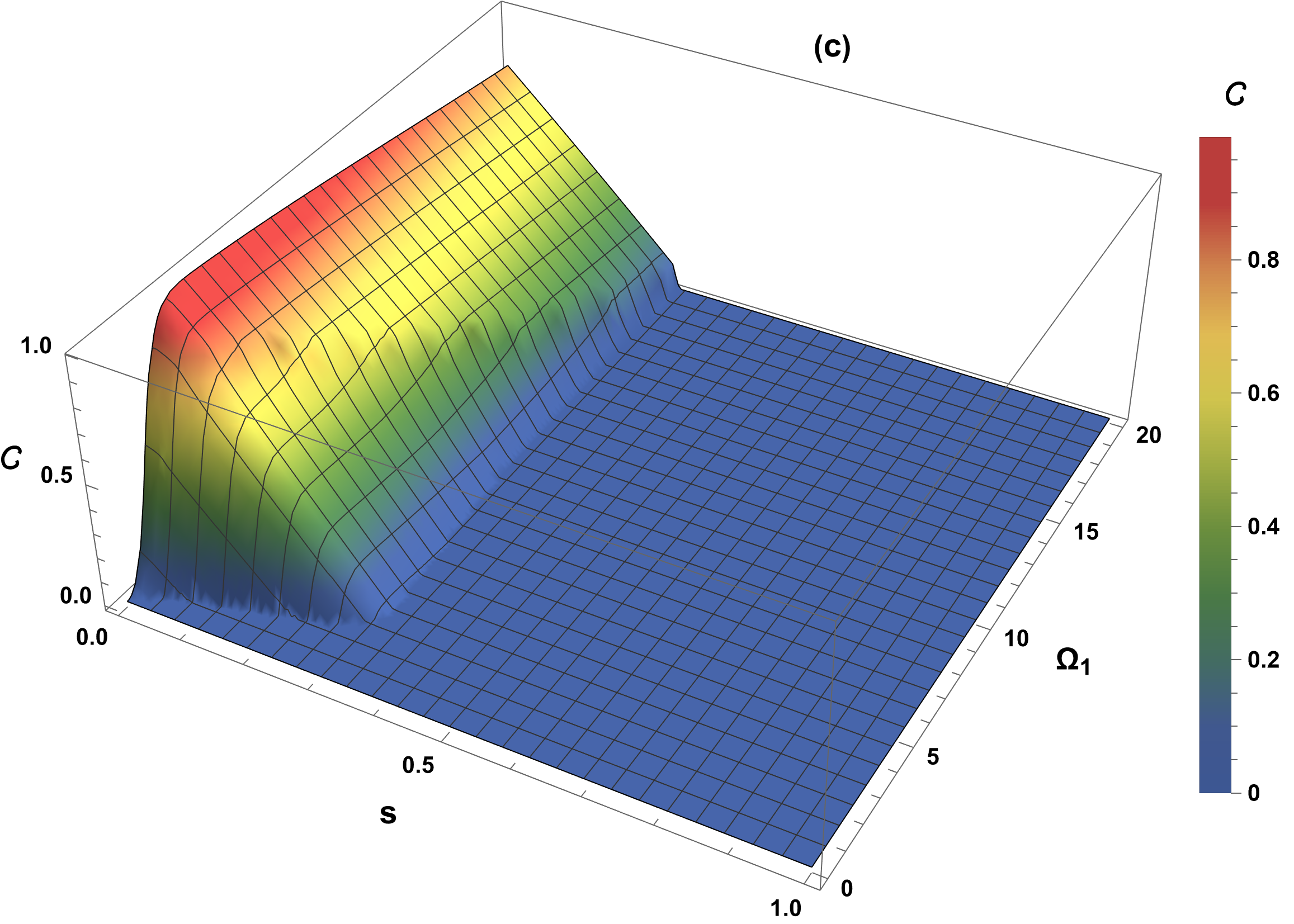}
\caption{
	Plot of the concurrence in the amplitude damping (AD) channel as a function of: 
	(a) the decoherence parameter $s$ and the temperature $\mathcal{T}$, with $\Omega_1 = 10$, $\Omega_2 = 15$, and $\mathcal{V} = 25$; 
	(b) the decoherence parameter $s$ and the Coulomb potential $\mathcal{V}$, with $\mathcal{T} = 0.1$, $\Omega_1 = 10$, and $\Omega_2 = 15$; 
	(c) the decoherence parameter $s$ and the coupling strength $\Omega_1 = \Omega_2$, with $\mathcal{T} = 0.1$ and $\mathcal{V} = 40$.
}
	\label{fig6}
\end{figure*}
Figure~\ref{fig6} displays the behavior of the concurrence $\mathcal{C}$ under the amplitude damping (AD) channel, which describes dissipative processes associated with energy loss.
The surface shown in Fig.~\ref{fig6}(a) reveals that entanglement is highly sensitive to both the decoherence parameter $s$ and the temperature $\mathcal{T}$. For small values of $s$, the concurrence remains significant, especially at low temperatures. However, as $s$ increases, $\mathcal{C}$ rapidly decreases and eventually vanishes, indicating a strong suppression of entanglement due to dissipation. This effect is further enhanced at higher temperatures, where thermal fluctuations accelerate the loss of quantum correlations.
The influence of the Coulomb interaction $\mathcal{V}$, illustrated in Fig.~\ref{fig6}(b), shows that stronger interaction initially supports higher entanglement for small $s$. Nevertheless, as the decoherence parameter increases, the concurrence drops abruptly to zero regardless of the value of $\mathcal{V}$. This behavior highlights that, in the presence of amplitude damping, dissipation dominates over interaction effects, leading to an inevitable destruction of entanglement.
A similar trend is observed in Fig.~\ref{fig6}(c) when varying the tunneling coupling $\Omega_1=\Omega_2$. Although larger coupling enhances the initial amount of entanglement, it does not prevent its rapid decay as $s$ increases. The concurrence collapses beyond a critical value of the decoherence parameter, reflecting the irreversible nature of energy relaxation processes.
These results clearly indicate that the amplitude damping channel induces a rapid and irreversible degradation of entanglement, leading to its sudden disappearance. Unlike dephasing-type noise, dissipative dynamics strongly suppress quantum correlations, making entanglement particularly fragile under energy-loss mechanisms.
\begin{figure*}[t!]
	\includegraphics[width=0.33\linewidth]{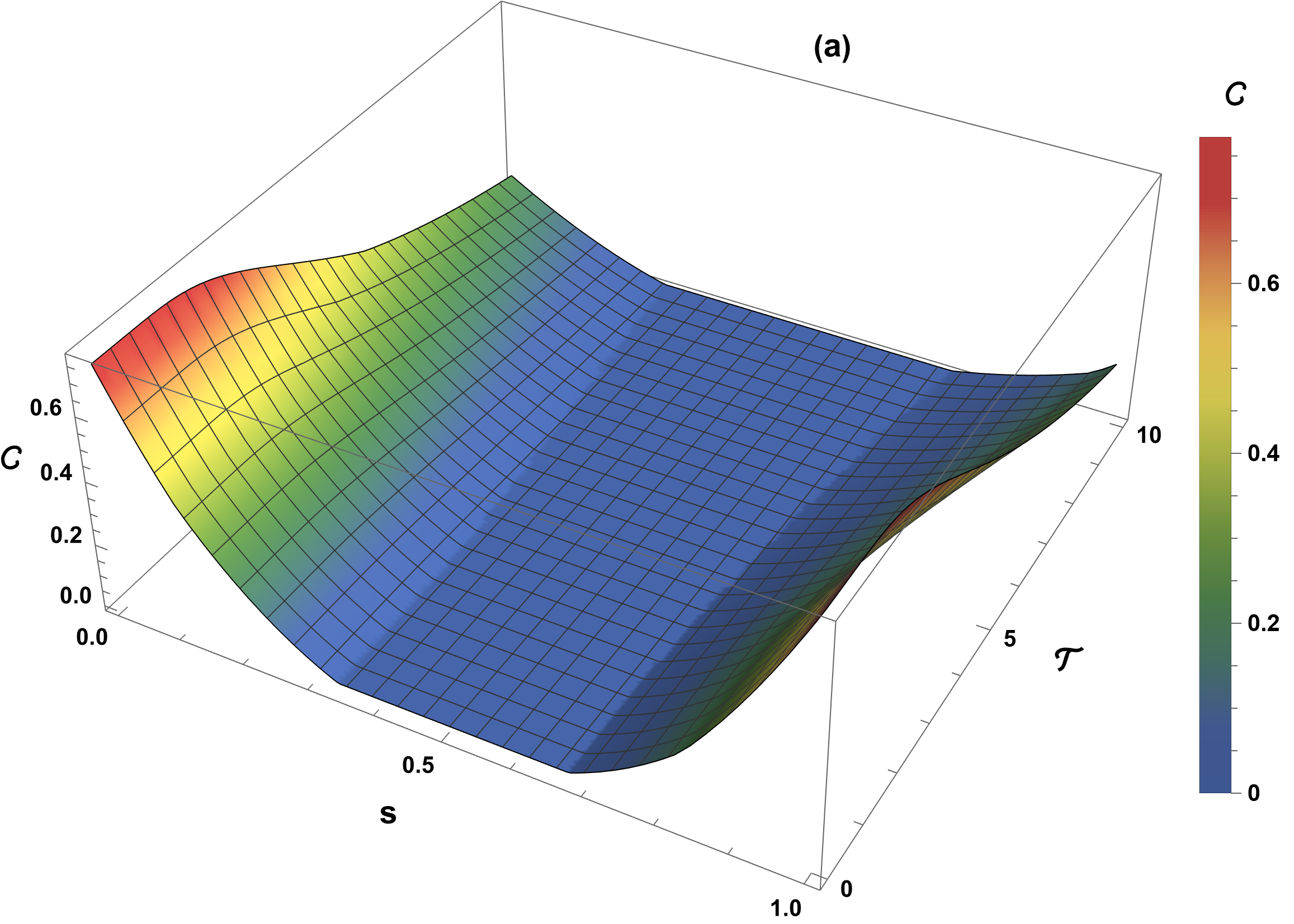}
	\includegraphics[width=0.33\linewidth]{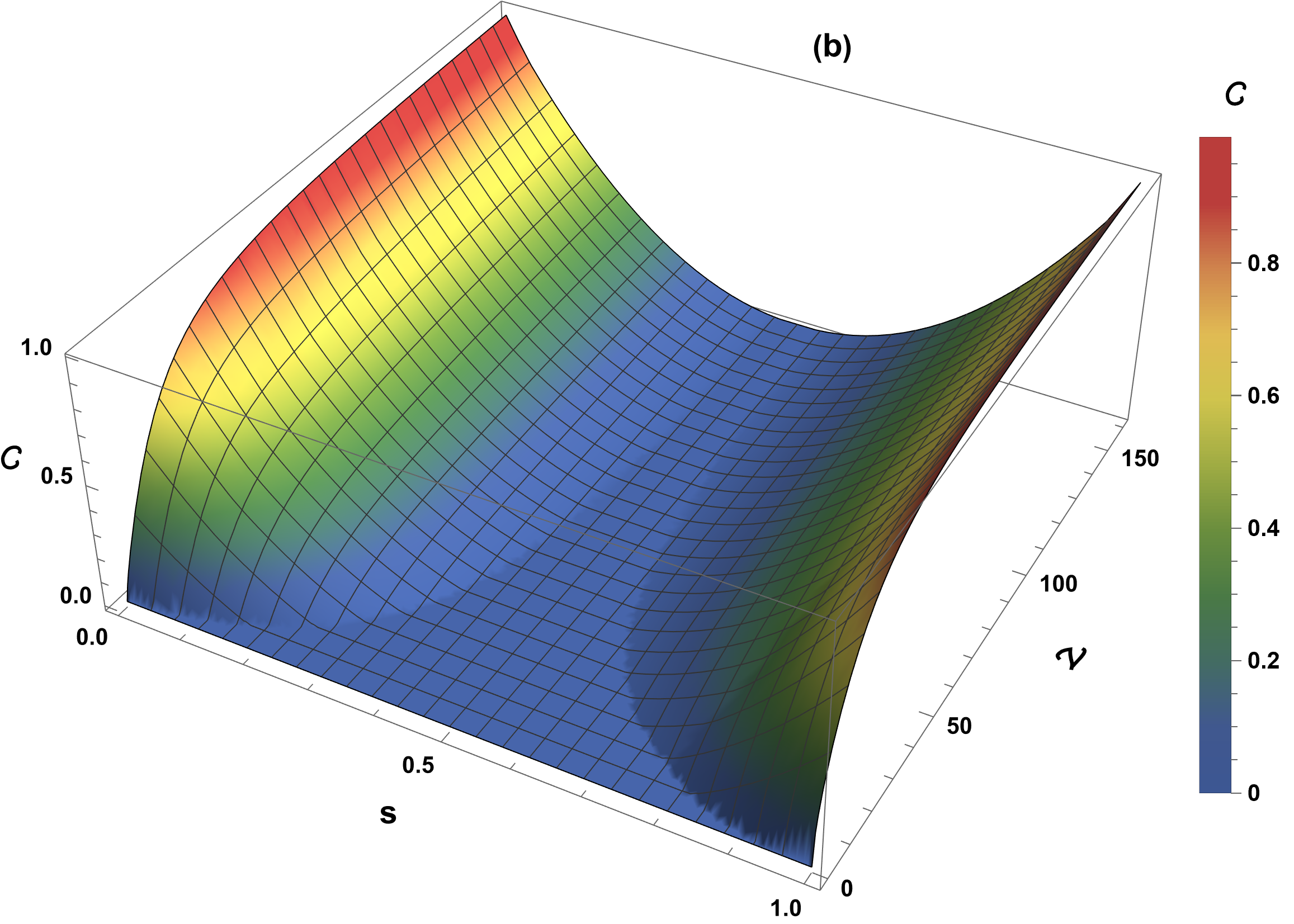}
	\includegraphics[width=0.33\linewidth]{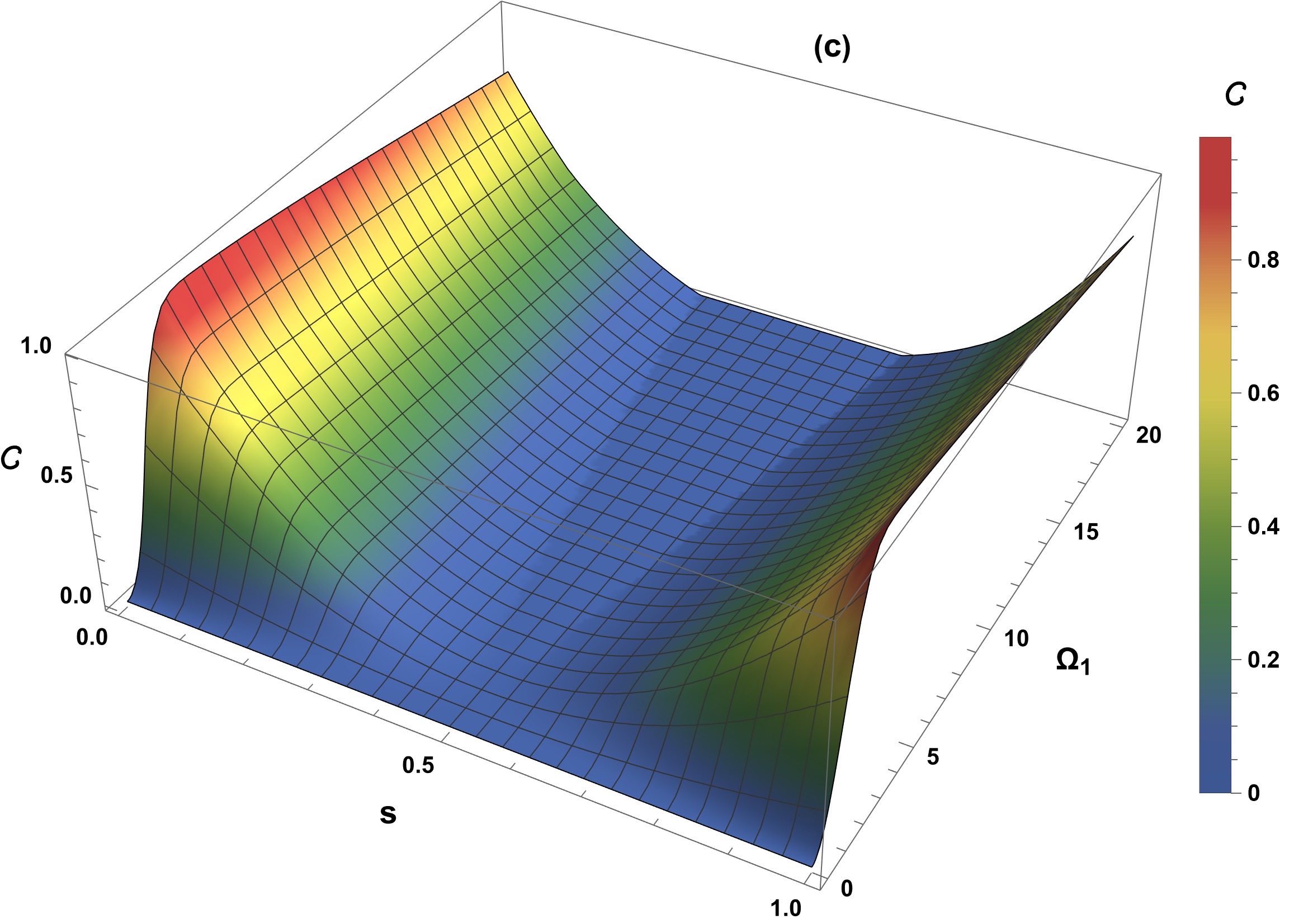}
\caption{
	Plot of the concurrence in the phase flip (PF) channel as a function of: 
	(a) the decoherence parameter $s$ and the temperature $\mathcal{T}$, with $\Omega_1 = 10$, $\Omega_2 = 15$, and $\mathcal{V} = 25$; 
	(b) the decoherence parameter $s$ and the Coulomb potential $\mathcal{V}$, with $\mathcal{T} = 0.1$, $\Omega_1 = 10$, and $\Omega_2 = 15$; 
	(c) the decoherence parameter $s$ and the coupling strength $\Omega_1 = \Omega_2$, with $\mathcal{T} = 0.1$ and $\mathcal{V} = 40$.
}
	\label{fig7}
\end{figure*}
\begin{figure*}[t!]
	\includegraphics[width=0.33\linewidth]{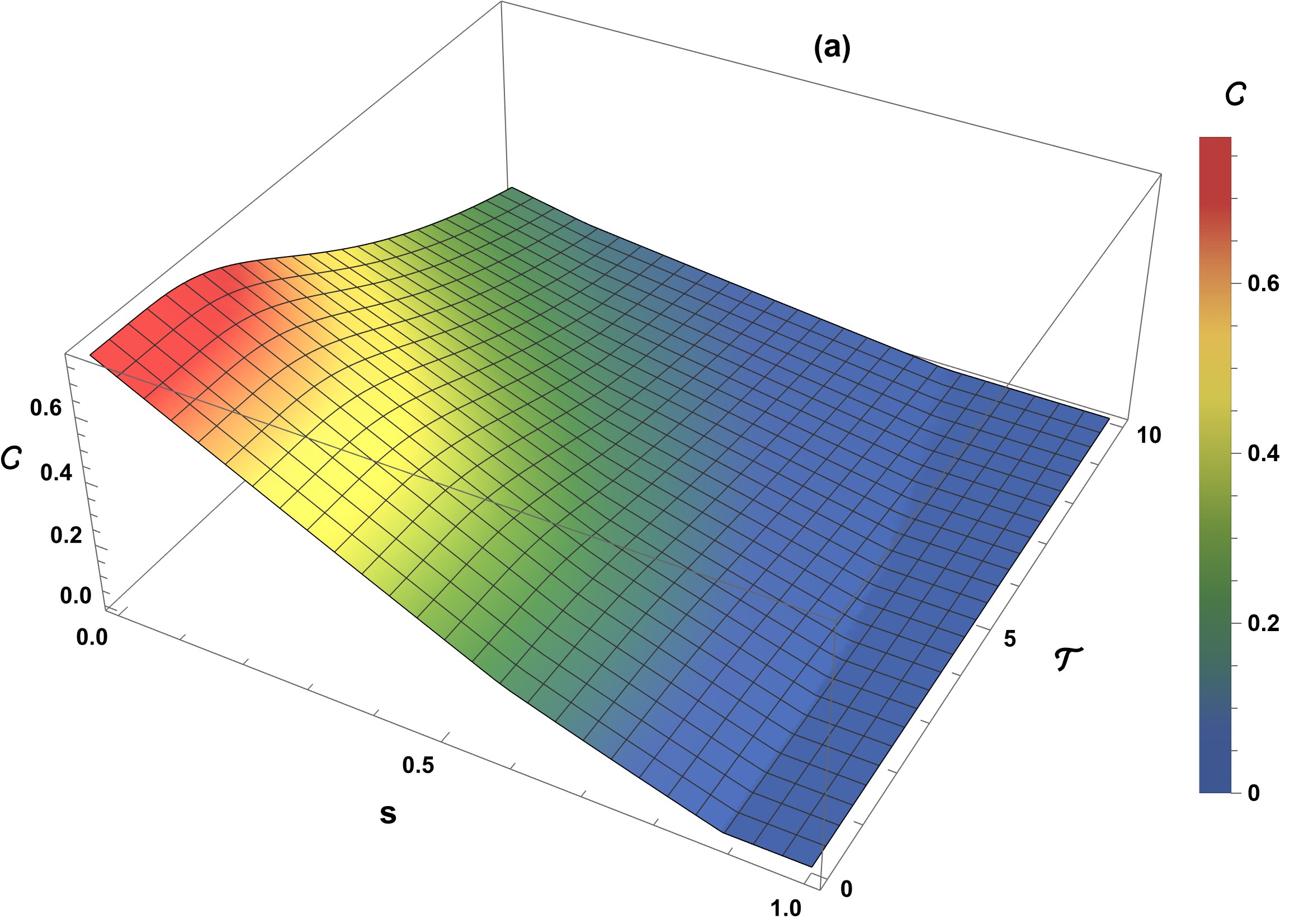}
	\includegraphics[width=0.33\linewidth]{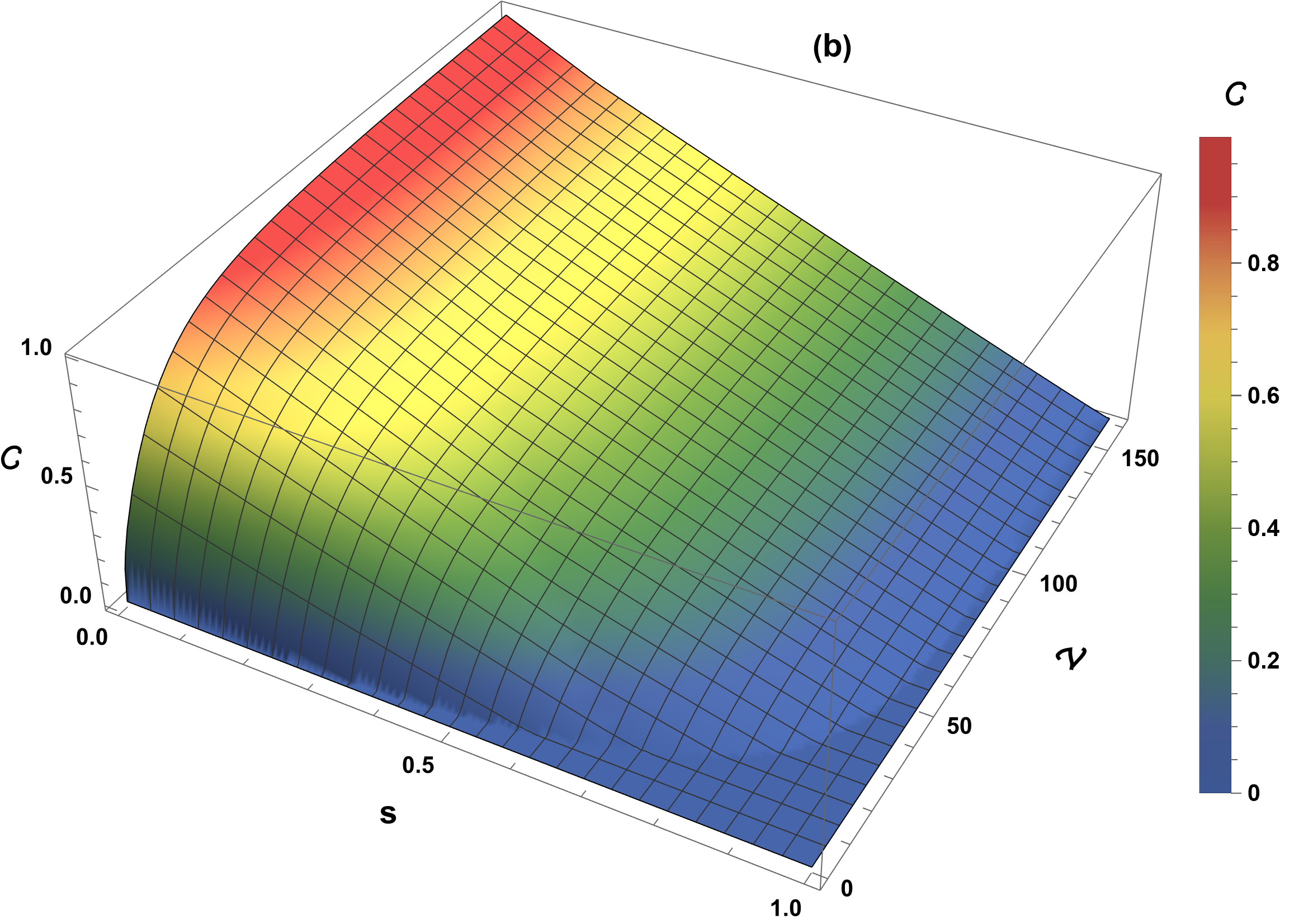}
	\includegraphics[width=0.33\linewidth]{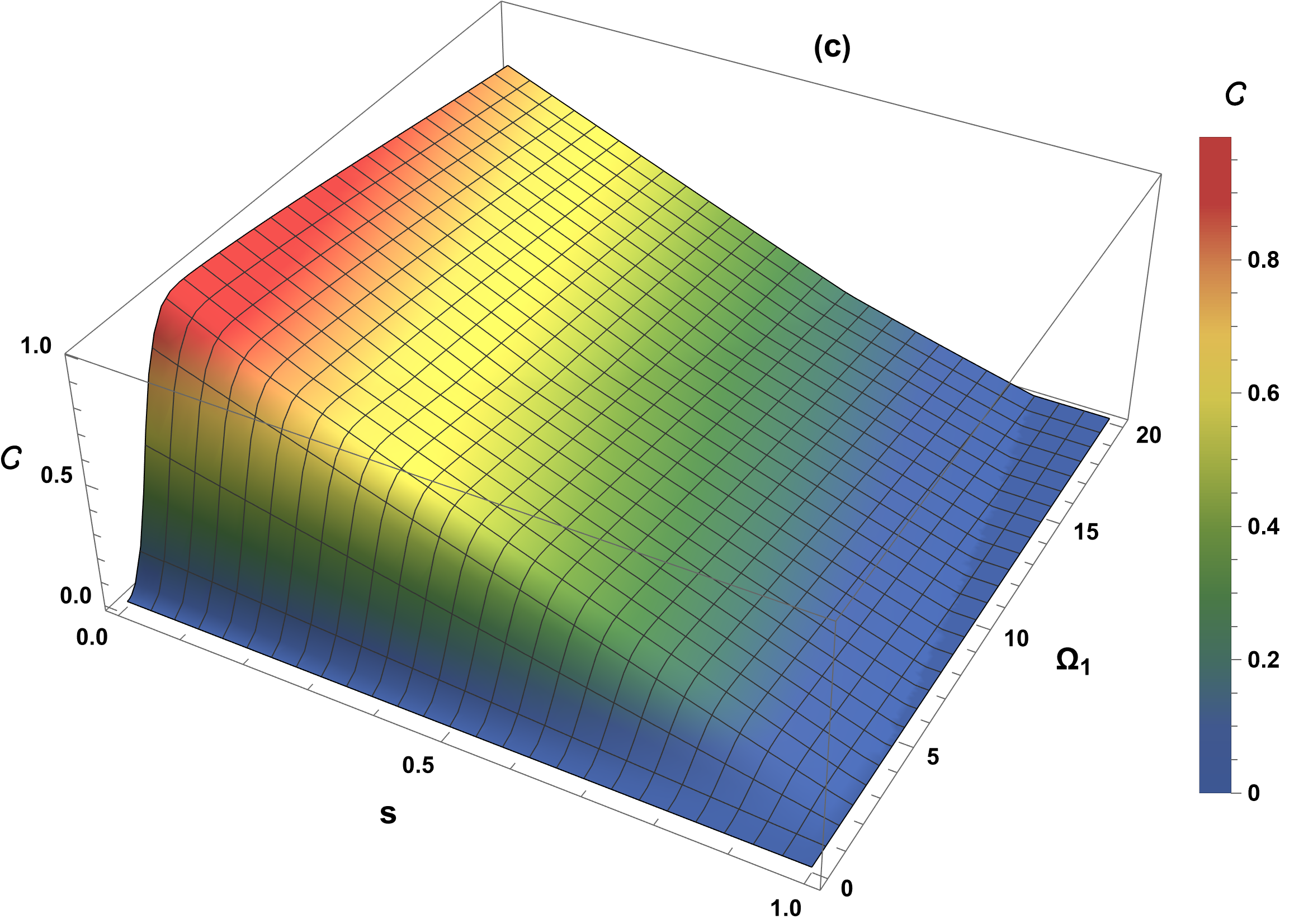}
	\caption{
		Plot of the concurrence in the phase damping (PD) channel as a function of: 
		(a) the decoherence parameter $s$ and the temperature $\mathcal{T}$, with $\Omega_1 = 10$, $\Omega_2 = 15$, and $\mathcal{V} = 25$; 
		(b) the decoherence parameter $s$ and the Coulomb potential $\mathcal{V}$, with $\mathcal{T} = 0.1$, $\Omega_1 = 10$, and $\Omega_2 = 15$; 
		(c) the decoherence parameter $s$ and the coupling strength $\Omega_1 = \Omega_2$, with $\mathcal{T} = 0.1$ and $\mathcal{V} = 40$.
	}
	\label{fig8}
\end{figure*}
\begin{figure*}[t!]
	\includegraphics[width=0.33\linewidth]{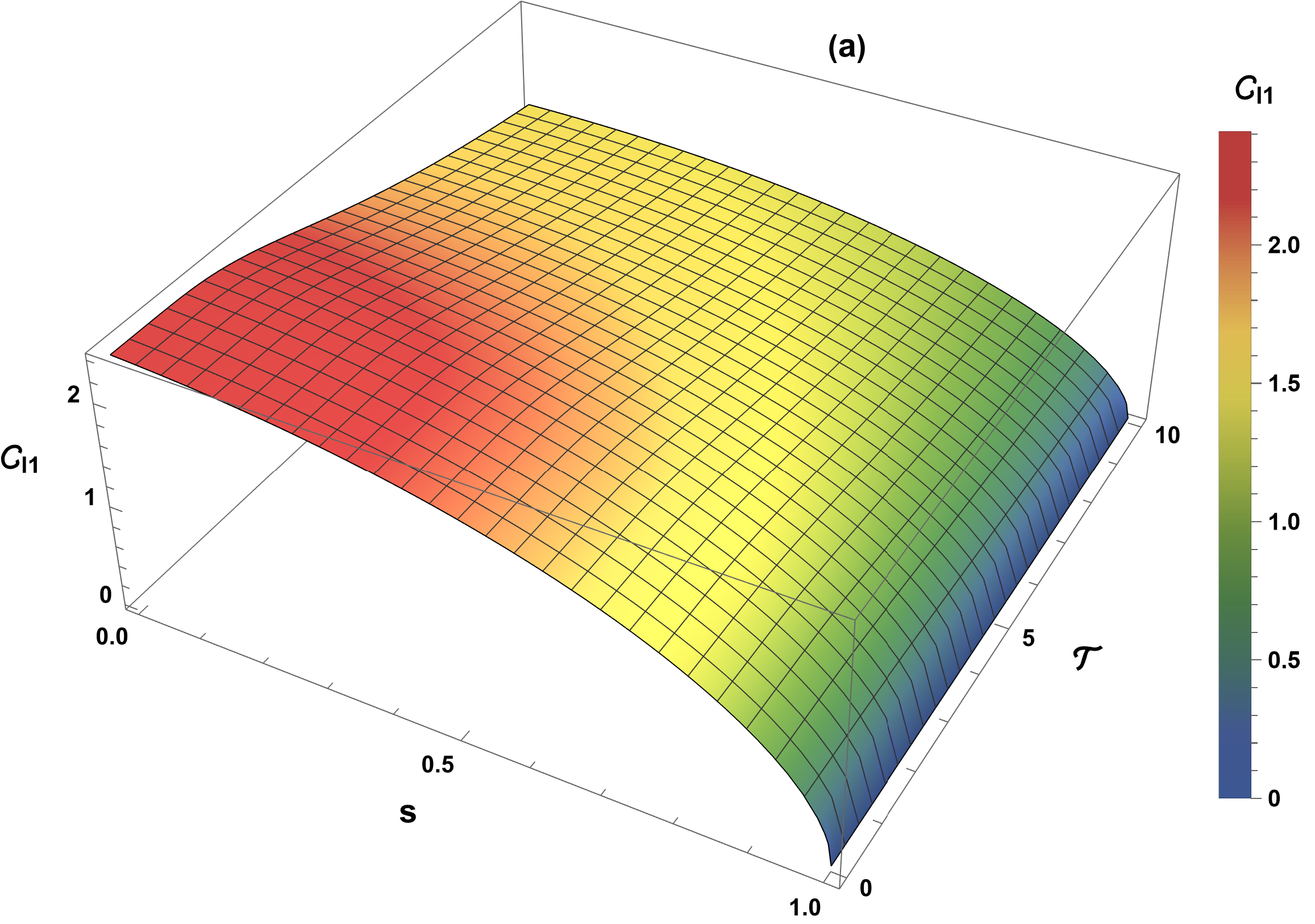}
	\includegraphics[width=0.33\linewidth]{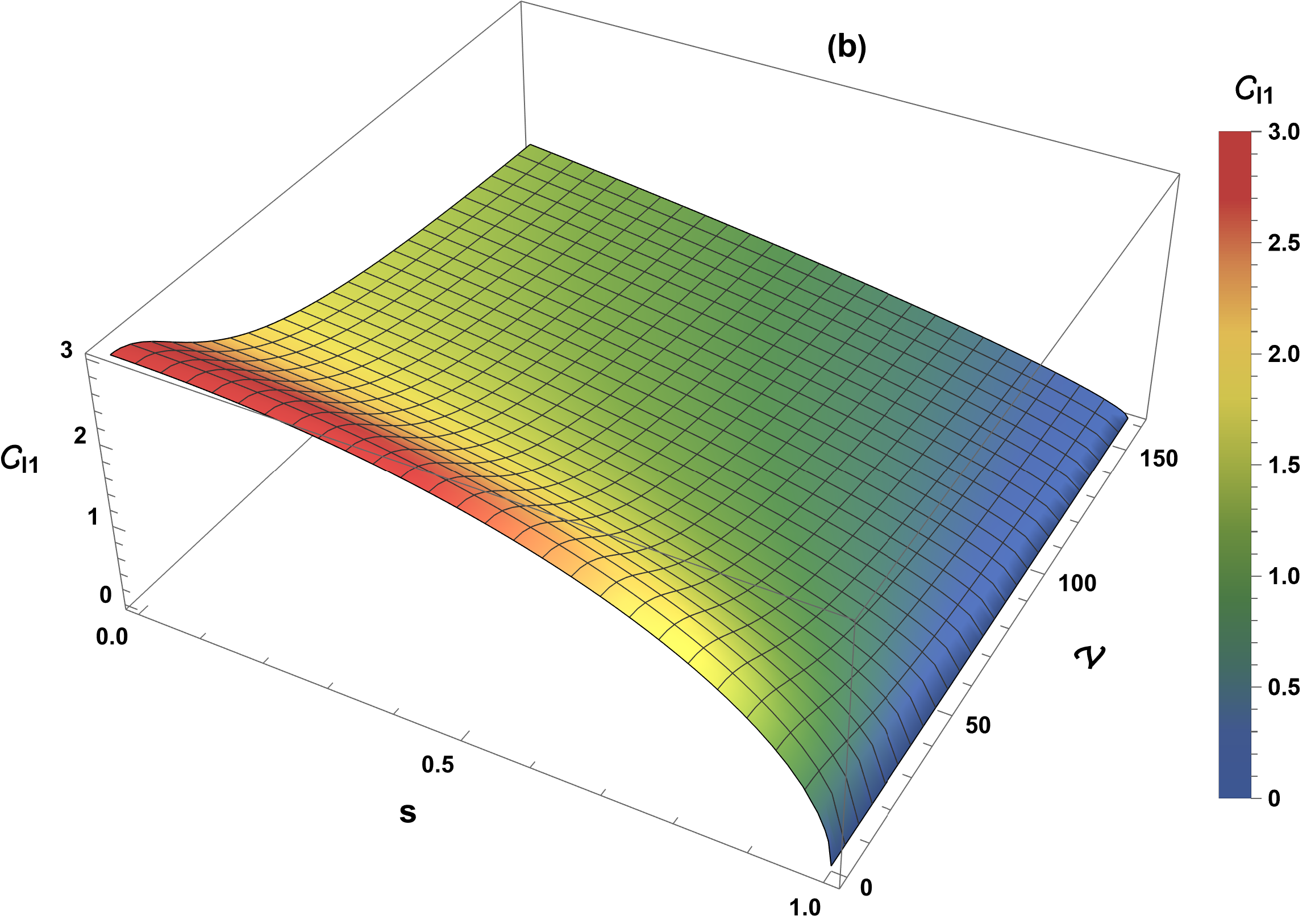}
	\includegraphics[width=0.33\linewidth]{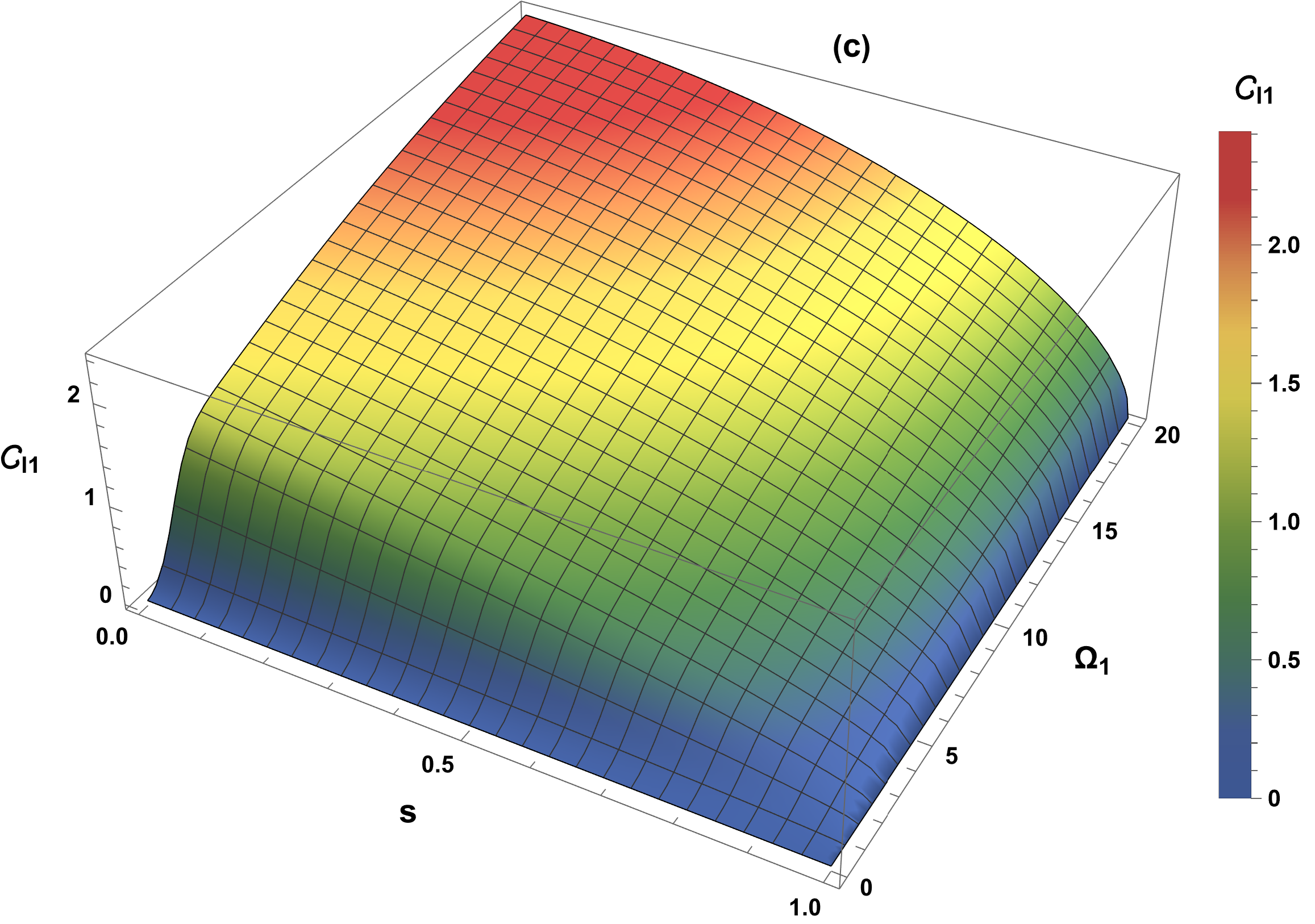}
	\caption{
		Plot of the quantum coherence in the amplitude damping (AD) channel as a function of: 
		(a) the decoherence parameter $s$ and the temperature $\mathcal{T}$, with $\Omega_1 = 10$, $\Omega_2 = 15$, and $\mathcal{V} = 25$; 
		(b) the decoherence parameter $s$ and the Coulomb potential $\mathcal{V}$, with $\mathcal{T} = 0.1$, $\Omega_1 = 10$, and $\Omega_2 = 15$; 
		(c) the decoherence parameter $s$ and the coupling strength $\Omega_1 = \Omega_2$, with $\mathcal{T} = 0.1$ and $\mathcal{V} = 40$.
	}
	\label{fig9}
\end{figure*}
\begin{figure*}[t!]
	\includegraphics[width=0.33\linewidth]{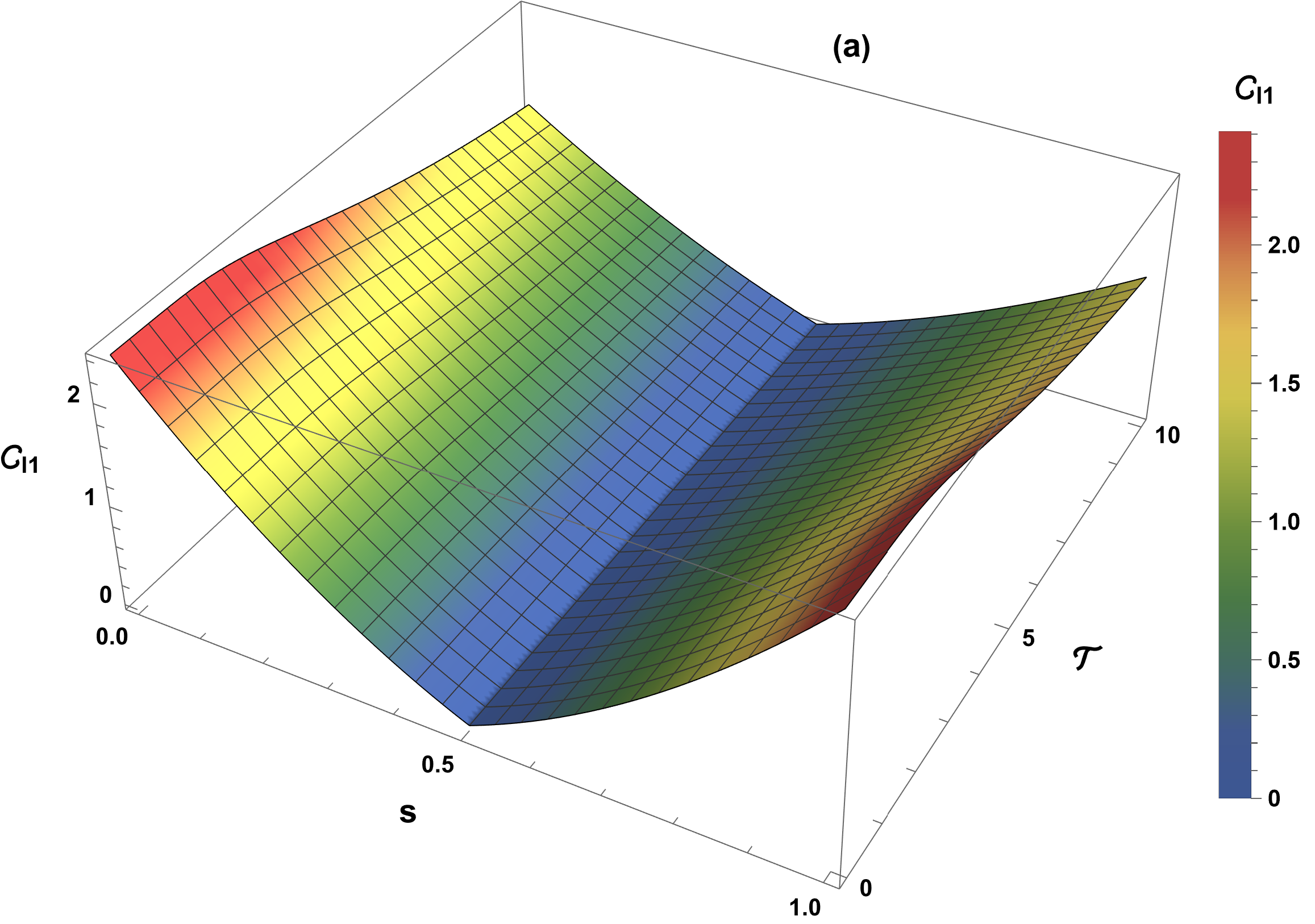}
	\includegraphics[width=0.33\linewidth]{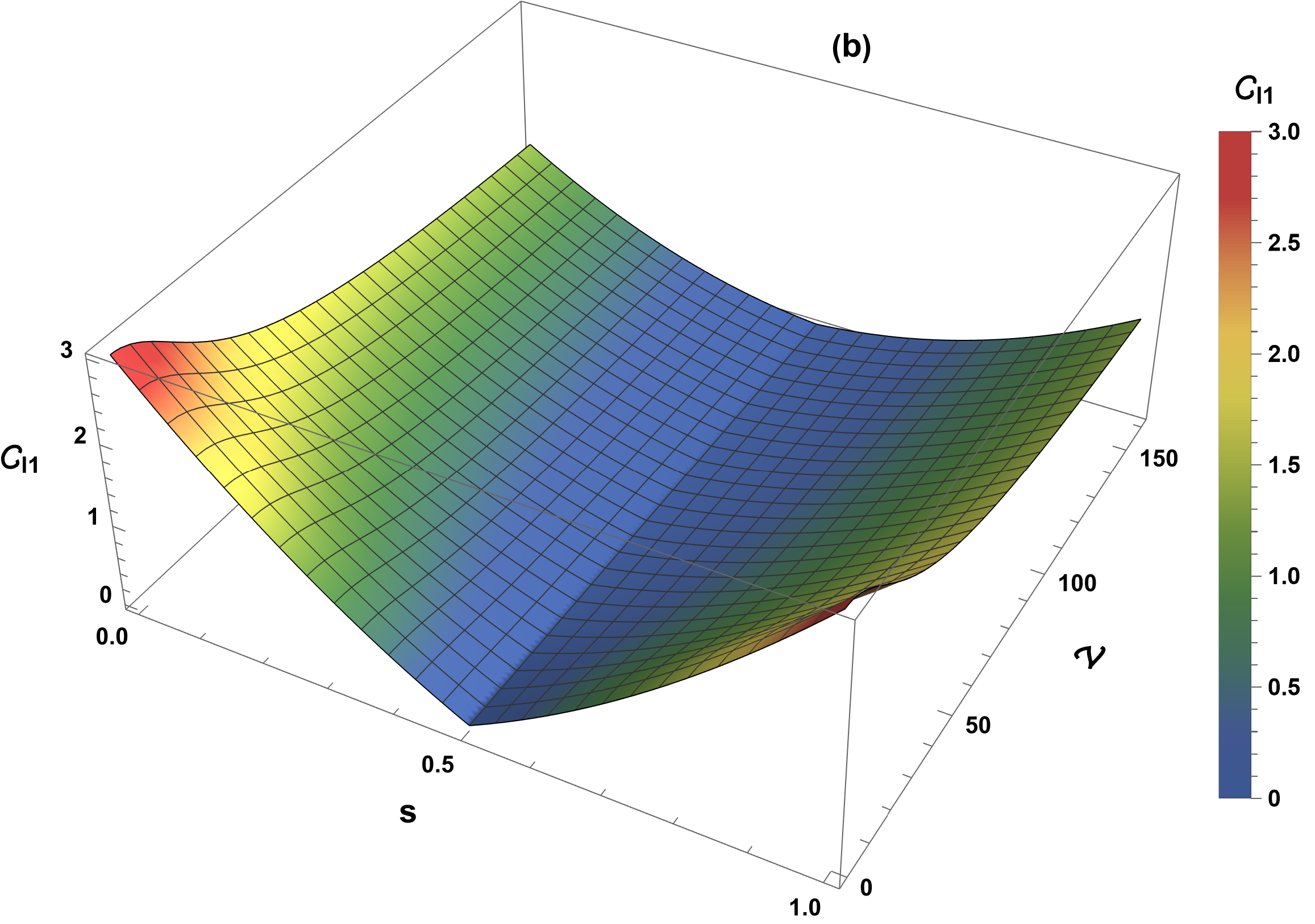}
	\includegraphics[width=0.33\linewidth]{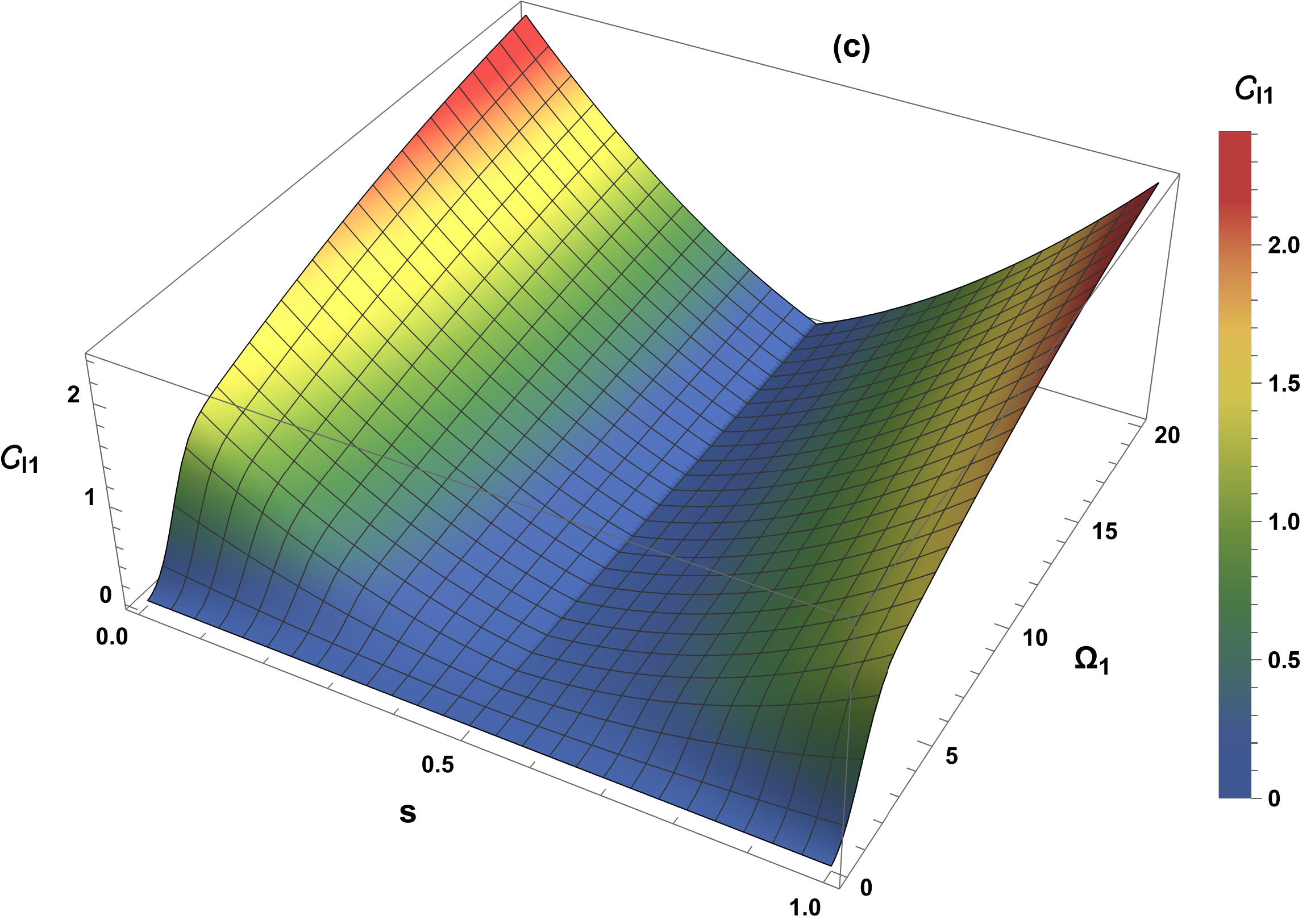}
	\caption{
		Plot of the quantum coherence in the phase flip (PF) channel as a function of: 
		(a) the decoherence parameter $s$ and the temperature $\mathcal{T}$, with $\Omega_1 = 10$, $\Omega_2 = 15$, and $\mathcal{V} = 25$; 
		(b) the decoherence parameter $s$ and the Coulomb potential $\mathcal{V}$, with $\mathcal{T} = 0.1$, $\Omega_1 = 10$, and $\Omega_2 = 15$; 
		(c) the decoherence parameter $s$ and the coupling strength $\Omega_1 = \Omega_2$, with $\mathcal{T} = 0.1$ and $\mathcal{V} = 40$.
	}
	\label{fig10}
\end{figure*}
\begin{figure*}[t!]
	\includegraphics[width=0.33\linewidth]{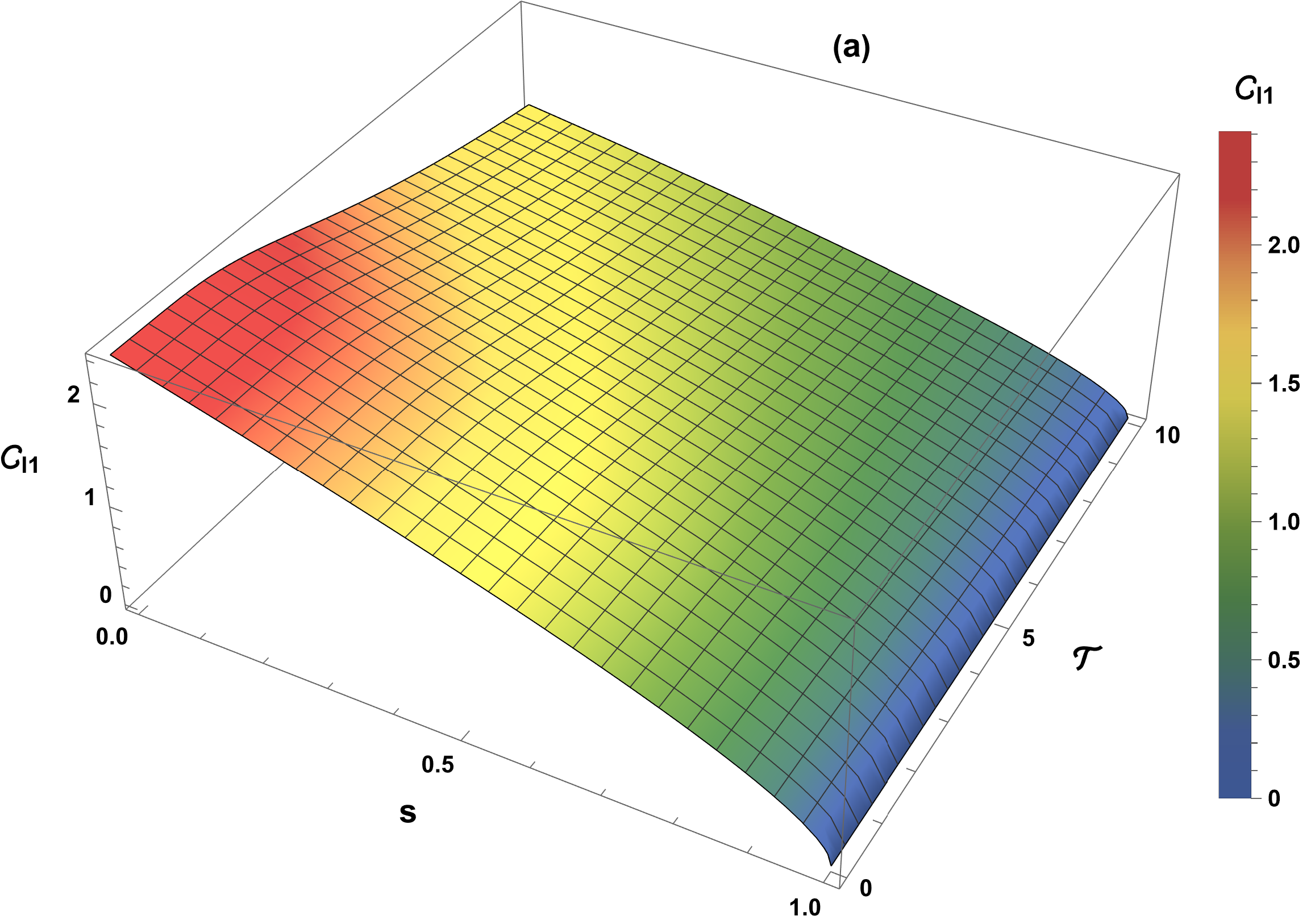}
	\includegraphics[width=0.33\linewidth]{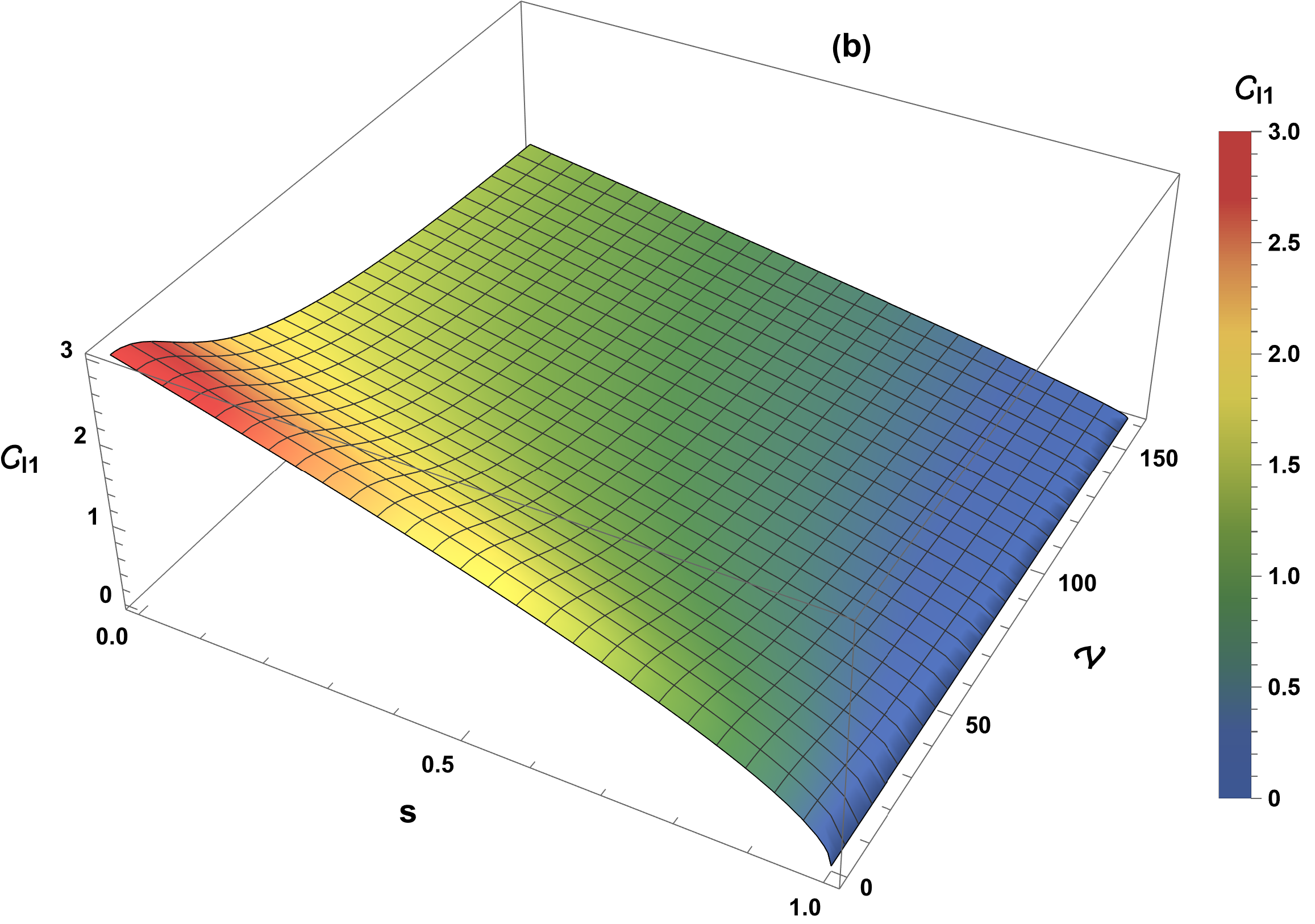}
	\includegraphics[width=0.33\linewidth]{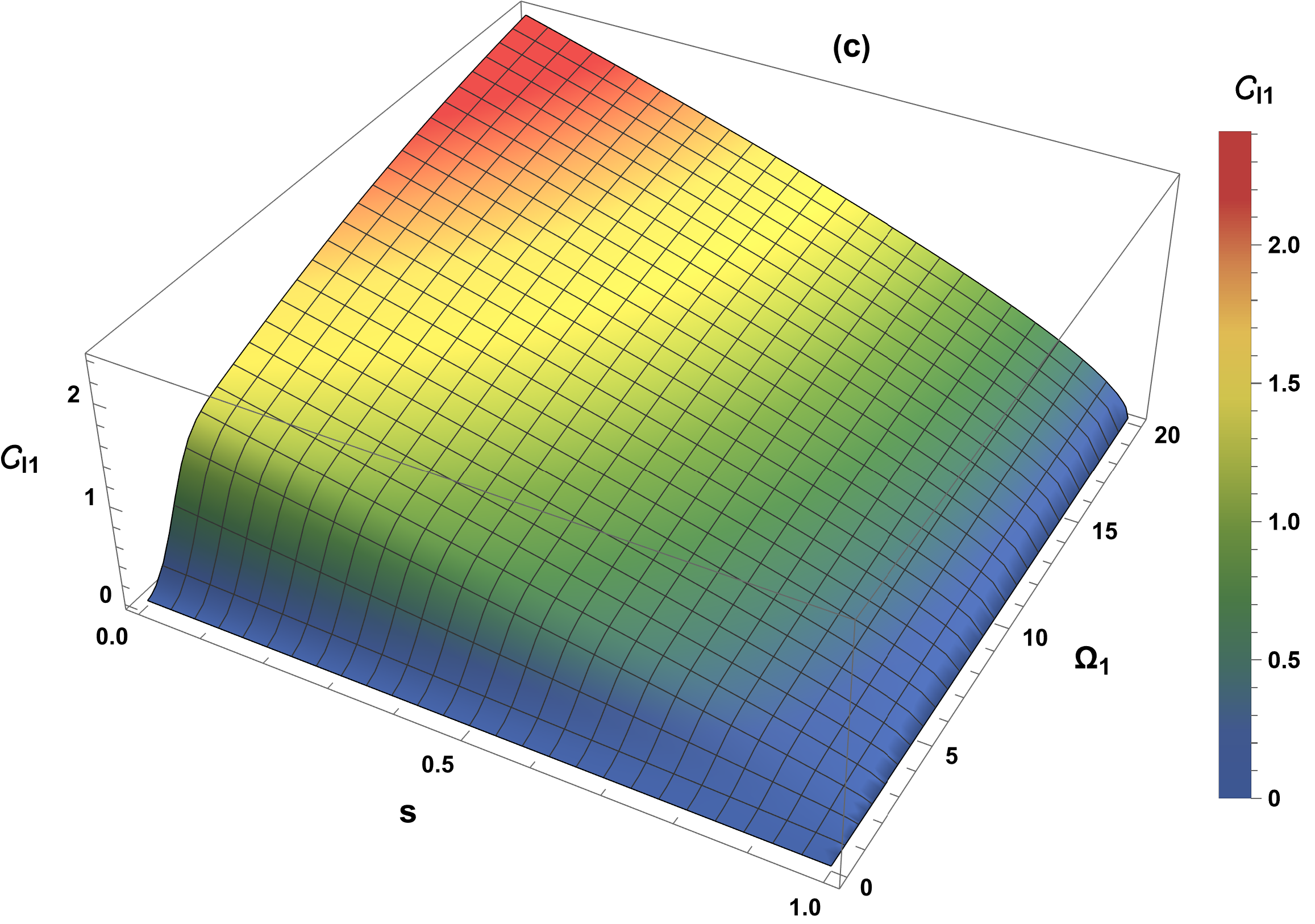}
	\caption{
		Plot of the quantum coherence in the phase damping (PD) channel as a function of: 
		(a) the decoherence parameter $s$ and the temperature $\mathcal{T}$, with $\Omega_1 = 10$, $\Omega_2 = 15$, and $\mathcal{V} = 25$; 
		(b) the decoherence parameter $s$ and the Coulomb potential $\mathcal{V}$, with $\mathcal{T} = 0.1$, $\Omega_1 = 10$, and $\Omega_2 = 15$; 
		(c) the decoherence parameter $s$ and the coupling strength $\Omega_1 = \Omega_2$, with $\mathcal{T} = 0.1$ and $\mathcal{V} = 40$.
	}
	\label{fig11}
\end{figure*}

Figure~\ref{fig7} illustrates the behavior of the concurrence $\mathcal{C}$ under the phase flip (PF) channel, which introduces phase errors without affecting the system energy.
A distinctive feature emerges in Fig.~\ref{fig7}(a), where the concurrence exhibits a non-monotonic dependence on the decoherence parameter $s$. Starting from a finite value at $s=0$, the entanglement decreases and reaches a minimum around intermediate values of $s$, before increasing again as $s$ approaches unity. This symmetric behavior originates from the nature of the PF channel, where the phase error becomes effectively equivalent to the identity operation at $s=1$. Temperature mainly affects the amplitude of concurrence, with higher temperatures leading to an overall reduction of entanglement.
The same qualitative structure is observed in Fig.~\ref{fig7}(b) when varying the Coulomb interaction $\mathcal{V}$. Larger values of $\mathcal{V}$ enhance the maximum achievable entanglement, but do not alter the characteristic non-monotonic profile as a function of $s$. This indicates that the PF channel preserves the underlying symmetry of the dynamics, while interaction strength controls only the magnitude of quantum correlations.
Figure~\ref{fig7}(c) shows that the tunneling coupling $\Omega_1=\Omega_2$ plays a similar role. Increasing the coupling leads to higher concurrence values, yet the global behavior remains unchanged, with a minimum at intermediate decoherence and recovery at large $s$. This confirms that the PF channel affects entanglement through phase disturbances rather than irreversible dissipation.
These results highlight a fundamental difference with the amplitude damping channel: in the PF case, entanglement is not irreversibly destroyed but instead undergoes a reversible-like degradation, with partial recovery at strong decoherence due to the intrinsic symmetry of the phase flip process.

Figure~\ref{fig8} captures the impact of pure dephasing on quantum correlations, where the loss of phase information governs the dynamics of entanglement.
In Fig.~\ref{fig8}(a), the effect of temperature is primarily reflected in the initial distribution of concurrence, while the subsequent evolution follows a uniform attenuation pattern as the decoherence parameter $s$ increases. This behavior highlights that phase damping acts by progressively erasing quantum superpositions rather than inducing abrupt transitions.
The dependence on the Coulomb interaction, shown in Fig.~\ref{fig8}(b), reveals that stronger interactions shift the concurrence to higher values without modifying the underlying decay mechanism. This indicates that interaction-induced correlations are present but remain vulnerable to phase noise, which steadily suppresses coherence between quantum states.
Figure~\ref{fig8}(c) emphasizes the role of tunneling in establishing the initial level of entanglement. Larger coupling strengths promote stronger quantum correlations, yet these are continuously degraded as dephasing accumulates, reflecting the inability of coherent dynamics to counterbalance phase noise.
Altogether, the PD channel imposes a uniform erosion of entanglement, governed by the gradual disappearance of off-diagonal contributions, and distinguishes itself by the absence of both sudden collapse and revival phenomena.

Figure~\ref{fig9} illustrates the behavior of quantum coherence $\mathcal{C}_{l_1}$ under the amplitude damping (AD) channel, where dissipation affects both populations and superposition terms.
From Fig.~\ref{fig9}(a), temperature mainly reshapes the initial distribution of coherence, while the evolution with respect to the decoherence parameter $s$ follows a continuous attenuation. Even for moderate values of $s$, coherence remains finite, indicating that superposition is only partially suppressed despite energy loss. The reduction becomes more pronounced at higher temperatures, where thermal mixing further weakens off-diagonal contributions.
The dependence on the Coulomb interaction, depicted in Fig.~\ref{fig9}(b), reveals a gradual reduction of coherence as $\mathcal{V}$ increases. This reflects the tendency of strong interactions to favor more localized configurations, thereby limiting the weight of coherent superpositions. Nevertheless, the decay induced by $s$ remains smooth, without abrupt suppression.
Figure~\ref{fig9}(c) shows that increasing the tunneling coupling $\Omega_1=\Omega_2$ enhances the overall level of coherence by promoting delocalized states. In this case, coherence spreads over a wider region of the parameter space, although it is still progressively damped as dissipation increases.
In contrast to entanglement, coherence under amplitude damping does not exhibit sudden disappearance but instead undergoes a gradual redistribution and attenuation, reflecting its intrinsic resilience as a measure of quantum superposition.

Figure~\ref{fig10} highlights the evolution of quantum coherence $\mathcal{C}_{l_1}$ under the phase flip (PF) channel, where decoherence acts exclusively on phase information without affecting populations.
A distinctive feature appears in Fig.~\ref{fig10}(a), where coherence exhibits a symmetric behavior with respect to the decoherence parameter $s$. Starting from its maximum value at $s=0$, it progressively decreases, reaches a minimum around intermediate values of $s$, and then increases again as $s \to 1$. This non-monotonic profile reflects the intrinsic nature of phase flip errors, which can partially reconstruct phase relations at strong noise levels. Temperature mainly controls the overall amplitude, with higher $\mathcal{T}$ leading to a global reduction of coherence.
The influence of the Coulomb interaction, shown in Fig.~\ref{fig10}(b), does not alter this symmetry but reshapes the landscape of coherence. Increasing $\mathcal{V}$ lowers the coherence in the intermediate region while preserving the recovery at large $s$, indicating that interaction-induced localization competes with phase-induced rephasing.
In Fig.~\ref{fig10}(c), the coupling strength $\Omega_1=\Omega_2$ plays a constructive role by elevating the coherence over the entire domain. Stronger coupling enhances the robustness of phase correlations, making the recovery at large $s$ more pronounced and extending the region where coherence remains significant.
Overall, the PF channel leads to a redistribution rather than a simple decay of coherence, governed by a balance between phase randomization and partial rephasing mechanisms.
\begin{figure*}[t!]
	\includegraphics[width=0.33\linewidth]{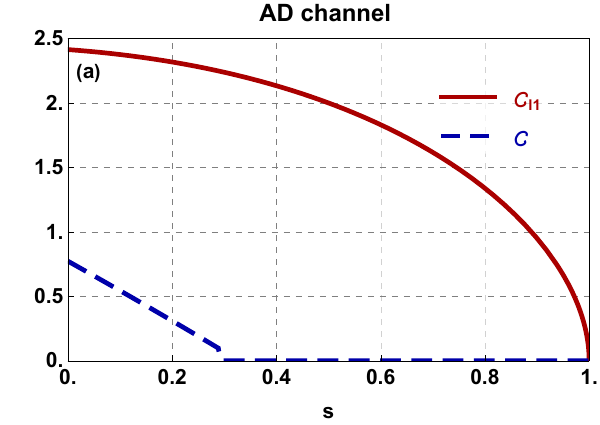}
	\includegraphics[width=0.33\linewidth]{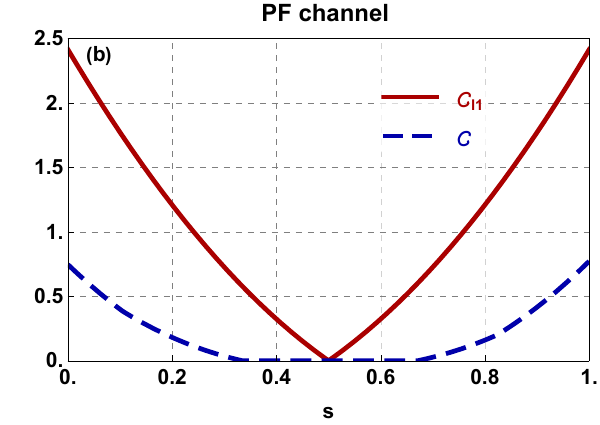}
	\includegraphics[width=0.33\linewidth]{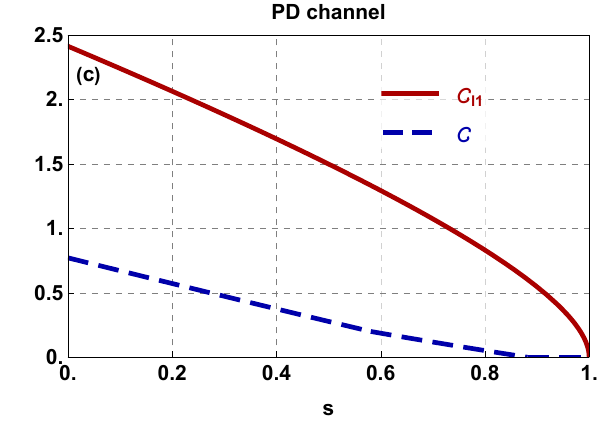}
	\caption{
		Comparison of the concurrence $\mathcal{C}$ and the quantum coherence $\mathcal{C}_{l_1}$ under different noisy channels: 
		(a) amplitude damping (AD); 
		(b) phase flip (PF); 
		(c) phase damping (PD). 
		The other parameters are set to $\mathcal{T} = 0.1$, $\mathcal{V} = 25$, $\Omega_1 = 10$, and $\Omega_2 = 15$.
	}
	\label{fig12}
\end{figure*}

Figure~\ref{fig11} illustrates the behavior of quantum coherence $\mathcal{C}_{l_1}$ under the phase damping (PD) channel, where decoherence affects exclusively the phase relations while leaving the populations unchanged.
As shown in Fig.~\ref{fig11}(a), coherence decreases smoothly and monotonically with the decoherence parameter $s$, without any revival or symmetry. This gradual attenuation reflects the nature of pure dephasing, which continuously destroys phase correlations. The effect of temperature is mainly quantitative, reducing the initial amount of coherence while preserving the same decay profile.
A similar tendency is observed in Fig.~\ref{fig11}(b), where increasing the Coulomb interaction $\mathcal{V}$ leads to a global suppression of coherence. Unlike the PF case, no restructuring or recovery occurs; instead, the interaction reinforces the loss of phase coherence throughout the entire evolution.
In Fig.~\ref{fig11}(c), the coupling strength $\Omega_1=\Omega_2$ enhances the initial coherence and slows down its degradation. Stronger coupling sustains phase correlations over a wider range of $s$, although it cannot prevent the inevitable decay induced by dephasing.
Overall, the PD channel induces a uniform and irreversible loss of coherence, governed by a purely diffusive mechanism that contrasts with both the dissipative nature of AD and the redistributive behavior of PF.
\newline
Figure~\ref{fig12} provides a direct comparison between the concurrence $\mathcal{C}$ and the quantum coherence $\mathcal{C}_{l_1}$ under different noisy channels, highlighting the distinct impact of each decoherence mechanism on quantum correlations.
In Fig.~\ref{fig12}(a), corresponding to the AD channel, both quantities exhibit a rapid decay with the decoherence parameter $s$, reflecting the dissipative nature of the environment. However, the concurrence vanishes faster, indicating that entanglement is more fragile than coherence under energy loss processes.
A contrasting behavior is observed in Fig.~\ref{fig12}(b) for the PF channel, where the concurrence displays a non-monotonic evolution, including a partial recovery at large values of $s$. In comparison, the coherence remains more stable and retains higher values throughout the evolution, emphasizing its robustness against phase-flip errors.
In Fig.~\ref{fig12}(c), associated with the PD channel, both concurrence and coherence decay monotonically, but with different sensitivities. The coherence decreases more gradually, while the concurrence is more strongly suppressed, confirming that pure dephasing predominantly affects quantum correlations without inducing any revival.
Taken together, these results clearly show that quantum coherence is generally more resilient than entanglement across all channels, while the qualitative dynamics strongly depend on the nature of the noise—dissipative for AD, redistributive for PF, and diffusive for PD.

In summary, the dynamics of quantum correlations, quantified here by the concurrence and the quantum coherence, strongly depends on the nature of the environmental interaction. While the amplitude damping channel leads to a rapid and irreversible degradation due to dissipative effects, the phase flip channel induces a redistribution of correlations with possible partial recovery. In contrast, the phase damping channel results in a smooth and monotonic decay governed by pure dephasing. Overall, quantum coherence remains more robust than concurrence across all channels.

\section{CONCLUSION}\label{sec8}

In this work, we have investigated the dynamics of quantum correlations, quantified by the concurrence and the quantum coherence, in a bipartite system under both Markovian and non-Markovian regimes, as well as in the presence of different noisy quantum channels. Our results reveal that the nature of the environment plays a decisive role in shaping the evolution of quantum correlations. In the Markovian regime, both entanglement and coherence undergo an irreversible decay, while non-Markovian effects induce memory-driven oscillations and partial revivals, reflecting a non-trivial backflow of information. When specific noise channels are considered, distinct mechanisms emerge: dissipative processes in the amplitude damping channel lead to a rapid suppression of correlations, whereas phase-based channels produce either a redistribution (phase flip) or a gradual degradation (phase damping). A central outcome of this study is the systematic robustness of quantum coherence compared to concurrence, highlighting its resilience against environmental disturbances and its potential as a reliable resource in realistic quantum systems. These findings provide deeper insight into the interplay between decoherence, interactions, and environmental memory, and may be relevant for the design of robust quantum information protocols. Future investigations could extend this analysis to correlated noise, higher-dimensional systems, or experimentally accessible platforms, in order to further bridge the gap between theoretical predictions and practical implementations.


\begin{thebibliography}{49}
	\makeatletter
	\providecommand \@ifxundefined [1]{%
		\@ifx{#1\undefined}
	}%
	\providecommand \@ifnum [1]{%
		\ifnum #1\expandafter \@firstoftwo
		\else \expandafter \@secondoftwo
		\fi
	}%
	\providecommand \@ifx [1]{%
		\ifx #1\expandafter \@firstoftwo
		\else \expandafter \@secondoftwo
		\fi
	}%
	\providecommand \natexlab [1]{#1}%
	\providecommand \enquote  [1]{``#1''}%
	\providecommand \bibnamefont  [1]{#1}%
	\providecommand \bibfnamefont [1]{#1}%
	\providecommand \citenamefont [1]{#1}%
	\providecommand \href@noop [0]{\@secondoftwo}%
	\providecommand \href [0]{\begingroup \@sanitize@url \@href}%
	\providecommand \@href[1]{\@@startlink{#1}\@@href}%
	\providecommand \@@href[1]{\endgroup#1\@@endlink}%
	\providecommand \@sanitize@url [0]{\catcode `\\12\catcode `\$12\catcode
		`\&12\catcode `\#12\catcode `\^12\catcode `\_12\catcode `\%12\relax}%
	\providecommand \@@startlink[1]{}%
	\providecommand \@@endlink[0]{}%
	\providecommand \url  [0]{\begingroup\@sanitize@url \@url }%
	\providecommand \@url [1]{\endgroup\@href {#1}{\urlprefix }}%
	\providecommand \urlprefix  [0]{URL }%
	\providecommand \Eprint [0]{\href }%
	\providecommand \doibase [0]{http://dx.doi.org/}%
	\providecommand \selectlanguage [0]{\@gobble}%
	\providecommand \bibinfo  [0]{\@secondoftwo}%
	\providecommand \bibfield  [0]{\@secondoftwo}%
	\providecommand \translation [1]{[#1]}%
	\providecommand \BibitemOpen [0]{}%
	\providecommand \bibitemStop [0]{}%
	\providecommand \EOS [0]{\spacefactor3000\relax}%
	\providecommand \BibitemShut  [1]{\csname bibitem#1\endcsname}%
	\let\auto@bib@innerbib\@empty
\bibitem [{\citenamefont {Einstein} \emph {et~al.}(1935)}]{Einstein1935}%
\BibitemOpen
\bibfield  {author} {\bibinfo {author} {\bibfnamefont {A.}~\bibnamefont {Einstein}}, \bibinfo {author} {\bibfnamefont {B.}~\bibnamefont {Podolsky}}, and \bibinfo {author} {\bibfnamefont {N.}~\bibnamefont {Rosen}},}
\bibfield  {title} {\enquote {\bibinfo {title} {Can Quantum-Mechanical Description of Physical Reality Be Considered Complete?},}}
\href {https://doi.org/10.1103/PhysRev.47.777}
{\bibfield  {journal} {\bibinfo  {journal} {Phys. Rev.}}
	\textbf {\bibinfo {volume} {47}},\
	\bibinfo {pages} {777} (\bibinfo {year} {1935})}
\BibitemShut {}%

\bibitem [{\citenamefont {Bell}(1964)}]{Bell1964}%
\BibitemOpen
\bibfield  {author} {\bibinfo {author} {\bibfnamefont {J.~S.}~\bibnamefont {Bell}},}
\bibfield  {title} {\enquote {\bibinfo {title} {On the Einstein Podolsky Rosen Paradox},}}
\href {https://doi.org/10.1103/PhysicsPhysiqueFizika.1.195}
{\bibfield  {journal} {\bibinfo  {journal} {Physics Physique Fizika}}
	\textbf {\bibinfo {volume} {1}},\
	\bibinfo {pages} {195} (\bibinfo {year} {1964})}
\BibitemShut {}%

\bibitem [{\citenamefont {Bennett et~al.}(1993)}]{Bennett1993}%
\BibitemOpen
\bibfield  {author} {\bibinfo {author} {\bibfnamefont {C.~H.}~\bibnamefont {Bennett}},
	\bibinfo {author} {\bibfnamefont {G.}~\bibnamefont {Brassard}},
	\bibinfo {author} {\bibfnamefont {C.}~\bibnamefont {Cr{\'e}peau}}, \textit{et al.},}
\bibfield  {title} {\enquote {\bibinfo {title} {Teleporting an Unknown Quantum State via Dual Classical and Einstein-Podolsky-Rosen Channels},}}
\href {https://doi.org/10.1103/PhysRevLett.70.1895}
{\bibfield  {journal} {\bibinfo  {journal} {Phys. Rev. Lett.}}
	\textbf {\bibinfo {volume} {70}},\
	\bibinfo {pages} {1895} (\bibinfo {year} {1993})}
\BibitemShut {}

\bibitem [{\citenamefont {Amico et~al.}(2008)}]{Amico2008}%
\BibitemOpen
\bibfield  {author} {\bibinfo {author} {\bibfnamefont {L.}~\bibnamefont {Amico}},
	\bibinfo {author} {\bibfnamefont {R.}~\bibnamefont {Fazio}},
	\bibinfo {author} {\bibfnamefont {A.}~\bibnamefont {Osterloh}},
	\bibinfo {author} {\bibfnamefont {V.}~\bibnamefont {Vedral}},}
\bibfield  {title} {\enquote {\bibinfo {title} {Entanglement in Many-Body Systems},}}
\href {https://doi.org/10.1103/RevModPhys.80.517}
{\bibfield  {journal} {\bibinfo  {journal} {Rev. Mod. Phys.}}
	\textbf {\bibinfo {volume} {80}},\
	\bibinfo {pages} {517} (\bibinfo {year} {2008})}
\BibitemShut {}

\bibitem [{\citenamefont {Schr\"odinger}(1935)}]{Schrodinger1935}%
\BibitemOpen
\bibfield  {author} {\bibinfo {author} {\bibfnamefont {E.}~\bibnamefont {Schr\"odinger}},}
\bibfield  {title} {\enquote {\bibinfo {title} {Discussion of probability relations between separated systems},}}
\href {https://doi.org/10.1017/S0305004100013554}
{\bibfield  {journal} {\bibinfo  {journal} {Math. Proc. Cambridge Philos. Soc.}}
	\textbf {\bibinfo {volume} {31}},\
	\bibinfo {pages} {555} (\bibinfo {year} {1935})}
\BibitemShut {}%

\bibitem [{\citenamefont {Benedetti} \emph {et~al.}(2007)}]{Benedetti2013}%
\BibitemOpen
\bibfield  {author} {\bibinfo {author} {\bibfnamefont {C.}~\bibnamefont {Benedetti}}, \bibinfo {author} {\bibfnamefont {F.}~\bibnamefont {Buscemi}}, \bibinfo {author} {\bibfnamefont {P.}~\bibnamefont {Bordone}}, and \bibinfo {author} {\bibfnamefont {M.~G.}~\bibnamefont {Paris}}}
\bibfield  {title} {\enquote {\bibinfo {title} {Dynamics of quantum correlations in colored-noise environments},}}
\href {https://doi.org/10.1103/PhysRevA.87.052328}
{\bibfield  {journal} {\bibinfo  {journal} {Phys. Rev. Lett.}}
	\textbf {\bibinfo {volume} {87}},\
	\bibinfo {pages} {052328} (\bibinfo {year} {2013})}
\BibitemShut {}%



\bibitem [{\citenamefont {Amazioug}(2023)}]{Amazioug2023}%
\BibitemOpen
\bibfield  {author} {\bibinfo {author} {\bibfnamefont {M.}~\bibnamefont {Amazioug}}, and \bibinfo {author} {\bibfnamefont {M.}~\bibnamefont {Daoud}},}
\bibfield  {title} {\enquote {\bibinfo {title} {Quantum steering vs entanglement and extracting work in an anisotropic two-qubit Heisenberg model in presence of external magnetic fields with DM and KSEA interactions},}}
\href {https://doi.org/10.1016/j.physleta.2023.129245}
{\bibfield  {journal} {\bibinfo  {journal} {Physics Letters A.}}
	\textbf {\bibinfo {volume} {493}},\
	\bibinfo {pages} {129245} (\bibinfo {year} {2024})}
\BibitemShut {}%


\bibitem [{\citenamefont {Jaloum}(2026)}]{Jaloum2026}%
\BibitemOpen
\bibfield  {author} {\bibinfo {author} {\bibfnamefont {E.}~\bibnamefont {Jaloum}}, and \bibinfo {author} {\bibfnamefont {M.}~\bibnamefont {Amazioug}},}
\bibfield  {title} {\enquote {\bibinfo {title} {Controlling the dynamical evolution of quantum coherence and quantum correlations in $e^+e^- -> \Lambda\bar{\Lambda}$ processes at BESIII},}}
\href {https://doi.org/10.1103/kxw9-wdth}
{\bibfield  {journal} {\bibinfo  {journal} {Physics Review D.}}
	\textbf {\bibinfo {volume} {113}},\
	\bibinfo {pages} {016024} (\bibinfo {year} {2026})}
\BibitemShut {}%


\bibitem [{\citenamefont {Jaloum}(2026)}]{JaloumNPB2026}%
\BibitemOpen
\bibfield  {author} {\bibinfo {author} {\bibfnamefont {E.}~\bibnamefont {Jaloum}}, and \bibinfo {author} {\bibfnamefont {M.}~\bibnamefont {Amazioug}},}
\bibfield  {title} {\enquote {\bibinfo {title} {Quantum teleportation, entanglement, LQU and LQFI in $e^+e^- -> \Lambda\bar{\Lambda}$ processes at BESIII through noisy channels},}}
\href {https://doi.org/10.1016/j.nuclphysb.2025.117255}
{\bibfield  {journal} {\bibinfo  {journal} {Nuclear Physics B.}}
	\textbf {\bibinfo {volume} {1022}},\
	\bibinfo {pages} {117255} (\bibinfo {year} {2026})}
\BibitemShut {}%


\bibitem [{\citenamefont {Wiseman} \emph {et~al.}(2007)}]{Wiseman2007}%
\BibitemOpen
\bibfield  {author} {\bibinfo {author} {\bibfnamefont {H.~M.}~\bibnamefont {Wiseman}}, \bibinfo {author} {\bibfnamefont {S.~J.}~\bibnamefont {Jones}}, and \bibinfo {author} {\bibfnamefont {A.~C.}~\bibnamefont {Doherty}},}
\bibfield  {title} {\enquote {\bibinfo {title} {Steering, Entanglement, Nonlocality, and the Einstein-Podolsky-Rosen Paradox},}}
\href {https://doi.org/10.1103/PhysRevLett.98.140402}
{\bibfield  {journal} {\bibinfo  {journal} {Phys. Rev. Lett.}}
	\textbf {\bibinfo {volume} {98}},\
	\bibinfo {pages} {140402} (\bibinfo {year} {2007})}
\BibitemShut {}%

\bibitem [{\citenamefont {Baumgratz et~al.}(2014)}]{Baumgratz2014}%
\BibitemOpen
\bibfield  {author} {\bibinfo {author} {\bibfnamefont {T.}~\bibnamefont {Baumgratz}},
	\bibinfo {author} {\bibfnamefont {M.}~\bibnamefont {Cramer}},
	\bibinfo {author} {\bibfnamefont {M.~B.}~\bibnamefont {Plenio}},}
\bibfield  {title} {\enquote {\bibinfo {title} {Quantifying Coherence},}}
\href {https://doi.org/10.1103/PhysRevLett.113.140401}
{\bibfield  {journal} {\bibinfo  {journal} {Phys. Rev. Lett.}}
	\textbf {\bibinfo {volume} {113}},\
	\bibinfo {pages} {140401} (\bibinfo {year} {2014})}
\BibitemShut {}

\bibitem [{\citenamefont {Streltsov et~al.}(2017)}]{Streltsov2017}%
\BibitemOpen
\bibfield  {author} {\bibinfo {author} {\bibfnamefont {A.}~\bibnamefont {Streltsov}},
	\bibinfo {author} {\bibfnamefont {G.}~\bibnamefont {Adesso}},
	\bibinfo {author} {\bibfnamefont {M.~B.}~\bibnamefont {Plenio}},}
\bibfield  {title} {\enquote {\bibinfo {title} {Colloquium: Quantum Coherence as a Resource},}}
\href {https://doi.org/10.1103/RevModPhys.89.041003}
{\bibfield  {journal} {\bibinfo  {journal} {Rev. Mod. Phys.}}
	\textbf {\bibinfo {volume} {89}},\
	\bibinfo {pages} {041003} (\bibinfo {year} {2017})}
\BibitemShut {}

\bibitem [{\citenamefont {Tan et~al.}(2016)}]{Tan2016}%
\BibitemOpen
\bibfield  {author} {\bibinfo {author} {\bibfnamefont {K.~C.}~\bibnamefont {Tan}},
	\bibinfo {author} {\bibfnamefont {H.}~\bibnamefont {Kwon}},
	\bibinfo {author} {\bibfnamefont {C.-Y.}~\bibnamefont {Park}},
	\bibinfo {author} {\bibfnamefont {H.}~\bibnamefont {Jeong}},}
\bibfield  {title} {\enquote {\bibinfo {title} {Unified View of Quantum Correlations and Quantum Coherence},}}
\href {https://doi.org/10.1103/PhysRevA.94.022329}
{\bibfield  {journal} {\bibinfo  {journal} {Phys. Rev. A}}
	\textbf {\bibinfo {volume} {94}},\
	\bibinfo {pages} {022329} (\bibinfo {year} {2016})}
\BibitemShut {}


\bibitem [{\citenamefont {Press} \emph {et~al.}(2008)}]{Press2008}%
\BibitemOpen
\bibfield  {author} {\bibinfo {author} {\bibfnamefont {D.}~\bibnamefont {Press}}, \bibinfo {author} {\bibfnamefont {T.~D.}~\bibnamefont {Ladd}}, \bibinfo {author} {\bibfnamefont {B.}~\bibnamefont {Zhang}}, and \bibinfo {author} {\bibfnamefont {Y.}~\bibnamefont {Yamamoto}},}
\bibfield  {title} {\enquote {\bibinfo {title} {Complete quantum control of a single quantum dot spin using ultrafast optical pulses},}}
\href {https://doi.org/10.1038/nature07530}
{\bibfield  {journal} {\bibinfo  {journal} {Nature}}
	\textbf {\bibinfo {volume} {456}},\
	\bibinfo {pages} {218} (\bibinfo {year} {2008})}
\BibitemShut {}%


\BibitemShut {}%
\bibitem [{\citenamefont {Economou}(2012)}]{Economou2012}%
\BibitemOpen
\bibfield  {author} {\bibinfo {author} {\bibfnamefont {M.~N.}~\bibnamefont {Economou}},}
\bibfield  {title} {\enquote {\bibinfo {title} {Quantum Mechanics: An Introduction},}}
\bibinfo {publisher} {Springer, Berlin} (\bibinfo {year} {2012})
\BibitemShut {}%

\bibitem [{\citenamefont {Loss and DiVincenzo}(1998)}]{Loss1998}%
\BibitemOpen
\bibfield  {author} {\bibinfo {author} {\bibfnamefont {D.}~\bibnamefont {Loss}},
	\bibinfo {author} {\bibfnamefont {D.~P.}~\bibnamefont {DiVincenzo}},}
\bibfield  {title} {\enquote {\bibinfo {title} {Quantum Computation with Quantum Dots},}}
\href {https://doi.org/10.1103/PhysRevA.57.120}
{\bibfield  {journal} {\bibinfo  {journal} {Phys. Rev. A}}
	\textbf {\bibinfo {volume} {57}},\
	\bibinfo {pages} {120} (\bibinfo {year} {1998})}
\BibitemShut {}

\bibitem [{\citenamefont {Gorman} \emph {et~al.}(2005)}]{Gorman2005}%
\BibitemOpen
\bibfield  {author} {\bibinfo {author} {\bibfnamefont {J.}~\bibnamefont {Gorman}}, \bibinfo {author} {\bibfnamefont {D.~G.}~\bibnamefont {Hasko}}, and \bibinfo {author} {\bibfnamefont {D.~A.}~\bibnamefont {Williams}},}
\bibfield  {title} {\enquote {\bibinfo {title} {Charge-Qubit Operation of an Isolated Double Quantum Dot},}}
\href {https://doi.org/10.1103/PhysRevLett.95.090502}
{\bibfield  {journal} {\bibinfo  {journal} {Phys. Rev. Lett.}}
	\textbf {\bibinfo {volume} {95}},\
	\bibinfo {pages} {090502} (\bibinfo {year} {2005})}
\BibitemShut {}%

\bibitem [{\citenamefont {Benito} \emph {et~al.}(2017)}]{Benito2017}%
\BibitemOpen
\bibfield  {author} {\bibinfo {author} {\bibfnamefont {M.}~\bibnamefont {Benito}}, \bibinfo {author} {\bibfnamefont {X.}~\bibnamefont {Mi}}, \bibinfo {author} {\bibfnamefont {J.~M.}~\bibnamefont {Taylor}}, \bibinfo {author} {\bibfnamefont {J.~R.}~\bibnamefont {Petta}}, and \bibinfo {author} {\bibfnamefont {G.}~\bibnamefont {Burkard}},}
\bibfield  {title} {\enquote {\bibinfo {title} {Input-output theory for spin-photon coupling in silicon double quantum dots},}}
\href {https://doi.org/10.1103/PhysRevB.96.235434}
{\bibfield  {journal} {\bibinfo  {journal} {Phys. Rev. B}}
	\textbf {\bibinfo {volume} {96}},\
	\bibinfo {pages} {235434} (\bibinfo {year} {2017})}
\BibitemShut {}%



\bibitem [{\citenamefont {D'Anjou} \emph {et~al.}(2019)}]{DAnjou2019}%
\BibitemOpen
\bibfield  {author} {\bibinfo {author} {\bibfnamefont {B.}~\bibnamefont {D'Anjou}} and \bibinfo {author} {\bibfnamefont {G.}~\bibnamefont {Burkard}},}
\bibfield  {title} {\enquote {\bibinfo {title} {Optimal exchange-based two-qubit gate for singlet-triplet qubits},}}
\href {https://doi.org/10.1103/PhysRevB.100.245427}
{\bibfield  {journal} {\bibinfo  {journal} {Phys. Rev. B}}
	\textbf {\bibinfo {volume} {100}},\
	\bibinfo {pages} {245427} (\bibinfo {year} {2019})}
\BibitemShut {}%

\bibitem [{\citenamefont {Shi} \emph {et~al.}(2012)}]{Shi2012}%
\BibitemOpen
\bibfield  {author} {\bibinfo {author} {\bibfnamefont {Z.}~\bibnamefont {Shi}}, \bibinfo {author} {\bibfnamefont {C.~B.}~\bibnamefont {Simmons}}, \bibinfo {author} {\bibfnamefont {J.~R.}~\bibnamefont {Prance}}, \bibinfo {author} {\bibfnamefont {J.~K.}~\bibnamefont {Gamble}}, \bibinfo {author} {\bibfnamefont {T.~S.}~\bibnamefont {Koh}}, \bibinfo {author} {\bibfnamefont {Y.-P.}~\bibnamefont {Shim}}, \bibinfo {author} {\bibfnamefont {X.}~\bibnamefont {Hu}}, \bibinfo {author} {\bibfnamefont {D.~E.}~\bibnamefont {Savage}}, \bibinfo {author} {\bibfnamefont {M.~G.}~\bibnamefont {Lagally}}, \bibinfo {author} {\bibfnamefont {M.~A.}~\bibnamefont {Eriksson}}, \bibinfo {author} {\bibfnamefont {M.}~\bibnamefont {Friesen}}, and \bibinfo {author} {\bibfnamefont {S.~N.}~\bibnamefont {Coppersmith}},}
\bibfield  {title} {\enquote {\bibinfo {title} {Coherent quantum oscillations and echo measurements of a Si charge qubit},}}
\href {https://doi.org/10.1103/PhysRevLett.108.140503}
{\bibfield  {journal} {\bibinfo  {journal} {Phys. Rev. Lett.}}
	\textbf {\bibinfo {volume} {108}},\
	\bibinfo {pages} {140503} (\bibinfo {year} {2012})}
\BibitemShut {}%

\bibitem [{\citenamefont {Yang} \emph {et~al.}(2020)}]{Yang2020}%
\BibitemOpen
\bibfield  {author} {\bibinfo {author} {\bibfnamefont {Y.-C.}~\bibnamefont {Yang}}, \bibinfo {author} {\bibfnamefont {S.~N.}~\bibnamefont {Coppersmith}}, and \bibinfo {author} {\bibfnamefont {M.}~\bibnamefont {Friesen}},}
\bibfield  {title} {\enquote {\bibinfo {title} {High-fidelity entangling gates in double-quantum-dot spin qubits},}}
\href {https://doi.org/10.1103/PhysRevA.101.012338}
{\bibfield  {journal} {\bibinfo  {journal} {Phys. Rev. A}}
	\textbf {\bibinfo {volume} {101}},\
	\bibinfo {pages} {012338} (\bibinfo {year} {2020})}
\BibitemShut {}%



\bibitem [{\citenamefont {Shinkai} \emph {et~al.}(2009)}]{Shinkai2009}%
\BibitemOpen
\bibfield  {author} {\bibinfo {author} {\bibfnamefont {G.}~\bibnamefont {Shinkai}}, \bibinfo {author} {\bibfnamefont {T.}~\bibnamefont {Hayashi}}, \bibinfo {author} {\bibfnamefont {T.}~\bibnamefont {Ota}}, and \bibinfo {author} {\bibfnamefont {T.}~\bibnamefont {Fujisawa}},}
\bibfield  {title} {\enquote {\bibinfo {title} {Correlated Coherent Oscillations in Coupled Semiconductor Charge Qubits},}}
\href {https://doi.org/10.1103/PhysRevLett.103.056802}
{\bibfield  {journal} {\bibinfo  {journal} {Phys. Rev. Lett.}}
	\textbf {\bibinfo {volume} {103}},\
	\bibinfo {pages} {056802} (\bibinfo {year} {2009})}
\BibitemShut {}%

\bibitem [{\citenamefont {Fujisawa} \emph {et~al.}(2004)}]{Fujisawa2004}%
\BibitemOpen
\bibfield  {author} {\bibinfo {author} {\bibfnamefont {T.}~\bibnamefont {Fujisawa}}, \bibinfo {author} {\bibfnamefont {T.}~\bibnamefont {Hayashi}}, and \bibinfo {author} {\bibfnamefont {Y.}~\bibnamefont {Hirayama}},}
\bibfield  {title} {\enquote {\bibinfo {title} {Time-dependent single-electron transport through quantum dots},}}
\href {https://doi.org/10.1116/1.1763894}
{\bibfield  {journal} {\bibinfo  {journal} {J. Vac. Sci. Technol. B}}
	\textbf {\bibinfo {volume} {22}},\
	\bibinfo {pages} {2035} (\bibinfo {year} {2004})}
\BibitemShut {}%

\bibitem [{\citenamefont {Austing} \emph {et~al.}(1998)}]{Austing1998}%
\BibitemOpen
\bibfield  {author} {\bibinfo {author} {\bibfnamefont {D.~G.}~\bibnamefont {Austing}}, \bibinfo {author} {\bibfnamefont {T.}~\bibnamefont {Honda}}, \bibinfo {author} {\bibfnamefont {K.}~\bibnamefont {Muraki}}, \bibinfo {author} {\bibfnamefont {Y.}~\bibnamefont {Tokura}}, and \bibinfo {author} {\bibfnamefont {S.}~\bibnamefont {Tarucha}},}
\bibfield  {title} {\enquote {\bibinfo {title} {Electronic states of semiconductor quantum dots},}}
{\bibfield  {journal} {\bibinfo  {journal} {Physica B}}
	\textbf {\bibinfo {volume} {249}},\
	\bibinfo {pages} {206} (\bibinfo {year} {1998})}
	
\bibitem [{\citenamefont {Filgueiras et~al.}(2020)}]{Filgueiras2020}%
\BibitemOpen
\bibfield  {author} {\bibinfo {author} {\bibfnamefont {J.~G.}~\bibnamefont {Filgueiras}},
	\bibinfo {author} {\bibfnamefont {B.~A.}~\bibnamefont {Castro}},
	\bibinfo {author} {\bibfnamefont {M.}~\bibnamefont {Sanz}},}
\bibfield  {title} {\enquote {\bibinfo {title} {Correlated Coherence in Thermal States of Quantum Systems},}}
\href {https://doi.org/10.1088/1367-2630/ab7b95}
{\bibfield  {journal} {\bibinfo  {journal} {New J. Phys.}}
	\textbf {\bibinfo {volume} {22}},\
	\bibinfo {pages} {033010} (\bibinfo {year} {2020})}
\BibitemShut {}

\bibitem [{\citenamefont {Filgueira} \emph {et~al.}(2020)}]{Filgueiras202}%
\BibitemOpen
\bibfield  {author} {\bibinfo {author} {\bibfnamefont {C.}~\bibnamefont {Filgueiras}}, \bibinfo {author} {\bibfnamefont {O.}~\bibnamefont {Rojas}}, and \bibinfo {author} {\bibfnamefont {M.}~\bibnamefont {Rojas}},}
\bibfield  {title} {\enquote {\bibinfo {title} {Thermal entanglement and correlated coherence in two coupled double quantum dots systems},}}
\href {https://doi.org/10.1002/andp.202000207}
{\bibfield  {journal} {\bibinfo  {journal} {Ann. Phys.}}
	\textbf {\bibinfo {volume} {532}},\
	\bibinfo {pages} {2000207} (\bibinfo {year} {2020})}
\BibitemShut {}%

\bibitem [{\citenamefont {Breuer and Petruccione}(2002)}]{BreuerPetruccione}%
\BibitemOpen
\bibfield  {author} {\bibinfo {author} {\bibfnamefont {H.-P.}~\bibnamefont {Breuer}},
	\bibinfo {author} {\bibfnamefont {F.}~\bibnamefont {Petruccione}},}
\bibfield  {title} {\enquote {\bibinfo {title} {The Theory of Open Quantum Systems},}}
{\bibfield  {journal} {\bibinfo  {journal} {Oxford University Press}}
	(\bibinfo {year} {2002})}
\BibitemShut {}

\bibitem [{\citenamefont {Khaetskii et~al.}(2002)}]{Khaetskii2002}%
\BibitemOpen
\bibfield  {author} {\bibinfo {author} {\bibfnamefont {A.~V.}~\bibnamefont {Khaetskii}},
	\bibinfo {author} {\bibfnamefont {D.}~\bibnamefont {Loss}},
	\bibinfo {author} {\bibfnamefont {L.}~\bibnamefont {Glazman}},}
\bibfield  {title} {\enquote {\bibinfo {title} {Electron Spin Decoherence in Quantum Dots},}}
\href {https://doi.org/10.1103/PhysRevLett.88.186802}
{\bibfield  {journal} {\bibinfo  {journal} {Phys. Rev. Lett.}}
	\textbf {\bibinfo {volume} {88}},\
	\bibinfo {pages} {186802} (\bibinfo {year} {2002})}
\BibitemShut {}

\bibitem [{\citenamefont {Coish and Loss}(2004)}]{Coish2004}%
\BibitemOpen
\bibfield  {author} {\bibinfo {author} {\bibfnamefont {W.~A.}~\bibnamefont {Coish}},
	\bibinfo {author} {\bibfnamefont {D.}~\bibnamefont {Loss}},}
\bibfield  {title} {\enquote {\bibinfo {title} {Hyperfine Interaction in a Quantum Dot},}}
\href {https://doi.org/10.1103/PhysRevB.70.195340}
{\bibfield  {journal} {\bibinfo  {journal} {Phys. Rev. B}}
	\textbf {\bibinfo {volume} {70}},\
	\bibinfo {pages} {195340} (\bibinfo {year} {2004})}
\BibitemShut {}

\bibitem [{\citenamefont {Petta et~al.}(2005)}]{Petta2005}%
\BibitemOpen
\bibfield  {author} {\bibinfo {author} {\bibfnamefont {J.~R.}~\bibnamefont {Petta}},}
\bibfield  {title} {\enquote {\bibinfo {title} {Coherent Manipulation of Coupled Electron Spins in Semiconductor Quantum Dots},}}
\href {https://doi.org/10.1126/science.1116955}
{\bibfield  {journal} {\bibinfo  {journal} {Science}}
	\textbf {\bibinfo {volume} {309}},\
	\bibinfo {pages} {2180} (\bibinfo {year} {2005})}
\BibitemShut {}

	

\bibitem [{\citenamefont {Yu} \emph {et~al.}(2004)}]{Yu2004}%
\BibitemOpen
\bibfield  {author} {\bibinfo {author} {\bibfnamefont {T.}~\bibnamefont {Yu}} and \bibinfo {author} {\bibfnamefont {J.~H.}~\bibnamefont {Eberly}},}
\bibfield  {title} {\enquote {\bibinfo {title} {Finite-Time Disentanglement Via Spontaneous Emission},}}
\href {https://doi.org/10.1103/PhysRevLett.93.140404}
{\bibfield  {journal} {\bibinfo  {journal} {Phys. Rev. Lett.}}
	\textbf {\bibinfo {volume} {93}},\
	\bibinfo {pages} {140404} (\bibinfo {year} {2004})}
\BibitemShut {}%

\bibitem [{\citenamefont {Palma} \emph {et~al.}(1996)}]{Palma1996}%
\BibitemOpen
\bibfield  {author} {\bibinfo {author} {\bibfnamefont {G.~M.}~\bibnamefont {Palma}}, \bibinfo {author} {\bibfnamefont {K.-A.}~\bibnamefont {Suominen}}, and \bibinfo {author} {\bibfnamefont {A.~K.}~\bibnamefont {Ekert}},}
\bibfield  {title} {\enquote {\bibinfo {title} {Quantum Computers and Dissipation},}}
\href {https://doi.org/10.1098/rspa.1996.0122}
{\bibfield  {journal} {\bibinfo  {journal} {Proc. R. Soc. A}}
	\textbf {\bibinfo {volume} {452}},\
	\bibinfo {pages} {567} (\bibinfo {year} {1996})}
\BibitemShut {}%

\bibitem [{\citenamefont {Breuer} \emph {et~al.}(2009)}]{Breuer2009}%
\BibitemOpen
\bibfield  {author} {\bibinfo {author} {\bibfnamefont {H.-P.}~\bibnamefont {Breuer}}, \bibinfo {author} {\bibfnamefont {E.-M.}~\bibnamefont {Laine}}, and \bibinfo {author} {\bibfnamefont {J.}~\bibnamefont {Piilo}},}
\bibfield  {title} {\enquote {\bibinfo {title} {Measure for the Degree of Non-Markovian Behavior of Quantum Processes},}}
\href {https://doi.org/10.1103/PhysRevLett.103.210401}
{\bibfield  {journal} {\bibinfo  {journal} {Phys. Rev. Lett.}}
	\textbf {\bibinfo {volume} {103}},\
	\bibinfo {pages} {210401} (\bibinfo {year} {2009})}
\BibitemShut {}%

\bibitem [{\citenamefont {Rivas} \emph {et~al.}(2010)}]{Rivas2010}%
\BibitemOpen
\bibfield  {author} {\bibinfo {author} {\bibfnamefont {Á.}~\bibnamefont {Rivas}}, \bibinfo {author} {\bibfnamefont {S.~F.}~\bibnamefont {Huelga}}, and \bibinfo {author} {\bibfnamefont {M.~B.}~\bibnamefont {Plenio}},}
\bibfield  {title} {\enquote {\bibinfo {title} {Entanglement and Non-Markovianity of Quantum Evolutions},}}
\href {https://doi.org/10.1103/PhysRevLett.105.050403}
{\bibfield  {journal} {\bibinfo  {journal} {Phys. Rev. Lett.}}
	\textbf {\bibinfo {volume} {105}},\
	\bibinfo {pages} {050403} (\bibinfo {year} {2010})}
\BibitemShut {}%

\bibitem [{\citenamefont {Breuer} \emph {et~al.}(2016)}]{BreuerReview2016}%
\BibitemOpen
\bibfield  {author} {\bibinfo {author} {\bibfnamefont {H.-P.}~\bibnamefont {Breuer}}, \bibinfo {author} {\bibfnamefont {E.-M.}~\bibnamefont {Laine}}, \bibinfo {author} {\bibfnamefont {J.}~\bibnamefont {Piilo}}, and \bibinfo {author} {\bibfnamefont {B.}~\bibnamefont {Vacchini}},}
\bibfield  {title} {\enquote {\bibinfo {title} {Colloquium: Non-Markovian dynamics in open quantum systems},}}
\href {https://doi.org/10.1103/RevModPhys.88.021002}
{\bibfield  {journal} {\bibinfo  {journal} {Rev. Mod. Phys.}}
	\textbf {\bibinfo {volume} {88}},\
	\bibinfo {pages} {021002} (\bibinfo {year} {2016})}
\BibitemShut {}%



\bibitem [{\citenamefont {Bachain} \emph {et~al.}(2026)}]{Bachain2026}%
\BibitemOpen
\bibfield  {author} {\bibinfo {author} {\bibfnamefont {O.}~\bibnamefont {Bachain}}, \bibinfo {author} {\bibfnamefont {M.}~\bibnamefont {Amazioug}}, \bibinfo {author} {\bibfnamefont {R.~A.}~\bibnamefont {Laamara}}, \bibinfo {author} {\bibfnamefont {K.~S.}~\bibnamefont {Nisar}}, \bibinfo {author} {\bibfnamefont {M.}~\bibnamefont {Zakarya}}, \bibinfo {author} {\bibfnamefont {G.~M.}~\bibnamefont {Ismail}}, and \bibinfo {author} {\bibfnamefont {A.-H.}~\bibnamefont {Abdel-Aty}},}
\bibfield  {title} {\enquote {\bibinfo {title} {Quantum thermodynamics, quantum correlations and quantum coherence in accelerating Unruh--DeWitt detectors in both steady and dynamical state},}}
\href {https://doi.org/10.1140/epjc/s10052-026-xxxxx}
{\bibfield  {journal} {\bibinfo  {journal} {Eur. Phys. J. C}}
	\textbf {\bibinfo {volume} {86}},\
	\bibinfo {pages} {271} (\bibinfo {year} {2026})}
\BibitemShut {}%

\bibitem [{\citenamefont {Eleuch} \emph {et~al.}(2017)}]{Eleuch2017}%
\BibitemOpen
\bibfield  {author} {\bibinfo {author} {\bibfnamefont {H.}~\bibnamefont {Eleuch}}, \bibinfo {author} {\bibfnamefont {M.}~\bibnamefont {Hilke}}, and \bibinfo {author} {\bibfnamefont {R.}~\bibnamefont {MacKenzie}},}
\bibfield  {title} {\enquote {\bibinfo {title} {Probing Anderson localization using the dynamics of a qubit},}}
\href {https://doi.org/10.1103/PhysRevA.95.062114}
{\bibfield  {journal} {\bibinfo  {journal} {Phys. Rev. A}}
	\textbf {\bibinfo {volume} {95}},\
	\bibinfo {pages} {062114} (\bibinfo {year} {2017})}
\BibitemShut {}%

\bibitem [{\citenamefont {Mohamed et~al.}(2022)}]{Mohamed2022}%
\BibitemOpen
\bibfield  {author} {\bibinfo {author} {\bibfnamefont {A.-B.~A.}~\bibnamefont {Mohamed}},
	\bibinfo {author} {\bibfnamefont {A.~U.}~\bibnamefont {Rahman}},
	\bibinfo {author} {\bibfnamefont {H.}~\bibnamefont {Eleuch}},}
\bibfield  {title} {\enquote {\bibinfo {title} {Measurement uncertainty, purity, and entanglement dynamics of maximally entangled two qubits interacting spatially with isolated cavities: intrinsic decoherence effect},}}
\href {https://doi.org/10.3390/e24040545}
{\bibfield  {journal} {\bibinfo  {journal} {Entropy}}
	\textbf {\bibinfo {volume} {24}},\
	\bibinfo {pages} {545} (\bibinfo {year} {2022})}
\BibitemShut {}
\bibitem [{\citenamefont {Fanchini} \emph {et~al.}(2010)}]{Fanchini2010}%
\BibitemOpen
\bibfield  {author} {\bibinfo {author} {\bibfnamefont {F.~F.}~\bibnamefont {Fanchini}}, \bibinfo {author} {\bibfnamefont {L.~K.}~\bibnamefont {Castelano}}, and \bibinfo {author} {\bibfnamefont {A.~O.}~\bibnamefont {Caldeira}},}
\bibfield  {title} {\enquote {\bibinfo {title} {Entanglement versus quantum discord in two coupled double quantum dots},}}
\href {https://doi.org/10.1088/1367-2630/12/7/073009}
{\bibfield  {journal} {\bibinfo  {journal} {New J. Phys.}}
	\textbf {\bibinfo {volume} {12}},\
	\bibinfo {pages} {073009} (\bibinfo {year} {2010})}
\BibitemShut {}%
\bibitem [{\citenamefont {Hu} \emph {et~al.}(2020)}]{Hu2020}%
\BibitemOpen
\bibfield  {author} {\bibinfo {author} {\bibfnamefont {M.-L.}~\bibnamefont {Hu}} and \bibinfo {author} {\bibfnamefont {H.}~\bibnamefont {Fan}},}
\bibfield  {title} {\enquote {\bibinfo {title} {Quantum coherence of multiqubit states in correlated noisy channels},}}
\href {https://doi.org/10.1007/s11433-019-1462-9}
{\bibfield  {journal} {\bibinfo  {journal} {Sci. China Phys. Mech. Astron.}}
	\textbf {\bibinfo {volume} {63}},\
	\bibinfo {pages} {230322} (\bibinfo {year} {2020})}
\BibitemShut {}%

\bibitem [{\citenamefont {Hu} \emph {et~al.}(2019)}]{Hu2019}%
\BibitemOpen
\bibfield  {author} {\bibinfo {author} {\bibfnamefont {M.}~\bibnamefont {Hu}} and \bibinfo {author} {\bibfnamefont {W.}~\bibnamefont {Zhou}},}
\bibfield  {title} {\enquote {\bibinfo {title} {Enhancing two-qubit quantum coherence in a correlated dephasing channel},}}
\href {https://doi.org/10.1088/1612-202X/ab0c6b}
{\bibfield  {journal} {\bibinfo  {journal} {Laser Phys. Lett.}}
	\textbf {\bibinfo {volume} {16}},\
	\bibinfo {pages} {045201} (\bibinfo {year} {2019})}
\BibitemShut {}%

\bibitem [{\citenamefont {Macchiavello} \emph {et~al.}(2002)}]{Macchiavello2002}%
\BibitemOpen
\bibfield  {author} {\bibinfo {author} {\bibfnamefont {C.}~\bibnamefont {Macchiavello}} and \bibinfo {author} {\bibfnamefont {G.~M.}~\bibnamefont {Palma}},}
\bibfield  {title} {\enquote {\bibinfo {title} {Entanglement-enhanced information transmission over a quantum channel with correlated noise},}}
\href {https://doi.org/10.1103/PhysRevA.65.050301}
{\bibfield  {journal} {\bibinfo  {journal} {Phys. Rev. A}}
	\textbf {\bibinfo {volume} {65}},\
	\bibinfo {pages} {050301} (\bibinfo {year} {2002})}
\BibitemShut {}%

\bibitem [{\citenamefont {Daffer} \emph {et~al.}(2004)}]{Daffer2004}%
\BibitemOpen
\bibfield  {author} {\bibinfo {author} {\bibfnamefont {S.}~\bibnamefont {Daffer}}, \bibinfo {author} {\bibfnamefont {K.}~\bibnamefont {W{\'o}dkiewicz}}, \bibinfo {author} {\bibfnamefont {J.~D.}~\bibnamefont {Cresser}}, and \bibinfo {author} {\bibfnamefont {J.~K.}~\bibnamefont {McIver}},}
\bibfield  {title} {\enquote {\bibinfo {title} {Depolarizing channel as a completely positive map with memory},}}
\href {https://doi.org/10.1103/PhysRevA.70.010304}
{\bibfield  {journal} {\bibinfo  {journal} {Phys. Rev. A}}
	\textbf {\bibinfo {volume} {70}},\
	\bibinfo {pages} {010304} (\bibinfo {year} {2004})}
\BibitemShut {}%

\bibitem [{\citenamefont {Nielsen} \emph {et~al.}(2010)}]{Nielsen2010}%
\BibitemOpen
\bibfield  {author} {\bibinfo {author} {\bibfnamefont {M.~A.}~\bibnamefont {Nielsen}} and \bibinfo {author} {\bibfnamefont {I.~L.}~\bibnamefont {Chuang}},}
\bibfield  {title} {\enquote {\bibinfo {title} {Quantum Computation and Quantum Information},}}
\bibinfo {publisher} {Cambridge University Press, Cambridge} (\bibinfo {year} {2010})
\BibitemShut {}%

\bibitem [{\citenamefont {Pirandola} \emph {et~al.}(2008)}]{Pirandola2008}%
\BibitemOpen
\bibfield  {author} {\bibinfo {author} {\bibfnamefont {S.}~\bibnamefont {Pirandola}}, \bibinfo {author} {\bibfnamefont {S.}~\bibnamefont {Mancini}}, \bibinfo {author} {\bibfnamefont {S.~L.}~\bibnamefont {Braunstein}}, and \bibinfo {author} {\bibfnamefont {D.}~\bibnamefont {Vitali}},}
\bibfield  {title} {\enquote {\bibinfo {title} {Minimal qudit code for a qubit in the phase-damping channel},}}
\href {https://doi.org/10.1103/PhysRevA.77.032309}
{\bibfield  {journal} {\bibinfo  {journal} {Phys. Rev. A}}
	\textbf {\bibinfo {volume} {77}},\
	\bibinfo {pages} {032309} (\bibinfo {year} {2008})}
\BibitemShut {}%

\bibitem [{\citenamefont {Damodarakurup} \emph {et~al.}(2009)}]{Damodarakurup2009}%
\BibitemOpen
\bibfield  {author} {\bibinfo {author} {\bibfnamefont {S.}~\bibnamefont {Damodarakurup}}, \bibinfo {author} {\bibfnamefont {M.}~\bibnamefont {Lucamarini}}, \bibinfo {author} {\bibfnamefont {G.}~\bibnamefont {Di~Giuseppe}}, \bibinfo {author} {\bibfnamefont {D.}~\bibnamefont {Vitali}}, and \bibinfo {author} {\bibfnamefont {P.}~\bibnamefont {Tombesi}},}
\bibfield  {title} {\enquote {\bibinfo {title} {Experimental inhibition of decoherence on flying qubits via ``bang-bang'' control},}}
\href {https://doi.org/10.1103/PhysRevLett.103.040502}
{\bibfield  {journal} {\bibinfo  {journal} {Phys. Rev. Lett.}}
	\textbf {\bibinfo {volume} {103}},\
	\bibinfo {pages} {040502} (\bibinfo {year} {2009})}
\BibitemShut {}%

\bibitem [{\citenamefont {Abd-Rabbou} \emph {et~al.}(2022)}]{AbdRabbou2022}%
\BibitemOpen
\bibfield  {author} {\bibinfo {author} {\bibfnamefont {M.~Y.}~\bibnamefont {Abd-Rabbou}}, \bibinfo {author} {\bibfnamefont {S.}~\bibnamefont {Khan}}, and \bibinfo {author} {\bibfnamefont {M.}~\bibnamefont {Shamirzaie}},}
\bibfield  {title} {\enquote {\bibinfo {title} {Quantum fisher information and quantum coherence of an entangled bipartite state interacting with a common classical environment in accelerating frames},}}
\href {https://doi.org/10.1007/s11128-022-03538-4}
{\bibfield  {journal} {\bibinfo  {journal} {Quantum Inf. Process.}}
	\textbf {\bibinfo {volume} {21}},\
	\bibinfo {pages} {218} (\bibinfo {year} {2022})}
\BibitemShut {}%
\bibitem [{\citenamefont {Wootters}(1998)}]{Wootters1998}%
\BibitemOpen
\bibfield  {author} {\bibinfo {author} {\bibfnamefont {W.~K.}~\bibnamefont {Wootters}},}
\bibfield  {title} {\enquote {\bibinfo {title} {Entanglement of formation of an arbitrary state of two qubits},}}
\href {https://doi.org/10.1103/PhysRevLett.80.2245}
{\bibfield  {journal} {\bibinfo  {journal} {Phys. Rev. Lett.}}
	\textbf {\bibinfo {volume} {80}},\
	\bibinfo {pages} {2245} (\bibinfo {year} {1998})}
\BibitemShut {}%

\bibitem [{\citenamefont {Hill} \emph {et~al.}(1997)}]{Hill1997}%
\BibitemOpen
\bibfield  {author} {\bibinfo {author} {\bibfnamefont {S.}~\bibnamefont {Hill}} and \bibinfo {author} {\bibfnamefont {W.~K.}~\bibnamefont {Wootters}},}
\bibfield  {title} {\enquote {\bibinfo {title} {Entanglement of a pair of quantum bits},}}
\href {https://doi.org/10.1103/PhysRevLett.78.5022}
{\bibfield  {journal} {\bibinfo  {journal} {Phys. Rev. Lett.}}
	\textbf {\bibinfo {volume} {78}},\
	\bibinfo {pages} {5022} (\bibinfo {year} {1997})}
\BibitemShut {}%

\bibitem [{\citenamefont {Nobel Committee}(2023)}]{NobelQD2023}%
\BibitemOpen
\bibfield  {author} {\bibinfo {author} {\bibnamefont {The Nobel Committee for Chemistry}},}
\bibfield  {title} {\enquote {\bibinfo {title} {The Nobel Prize in Chemistry 2023: Quantum dots},}}
\href {https://www.nobelprize.org/prizes/chemistry/2023/summary/}
{\bibfield  {journal} {\bibinfo  {journal} {NobelPrize.org}}
	(\bibinfo {year} {2023})}
\BibitemShut {}

\bibitem [{\citenamefont {Hanson et~al.}(2007)}]{Hanson2007}%
\BibitemOpen
\bibfield  {author} {\bibinfo {author} {\bibfnamefont {R.}~\bibnamefont {Hanson}},
	\bibinfo {author} {\bibfnamefont {L.~P.}~\bibnamefont {Kouwenhoven}},
	\bibinfo {author} {\bibfnamefont {J.~R.}~\bibnamefont {Petta}},
	\bibinfo {author} {\bibfnamefont {S.}~\bibnamefont {Tarucha}},
	\bibinfo {author} {\bibfnamefont {L.~M.~K.}~\bibnamefont {Vandersypen}},}
\bibfield  {title} {\enquote {\bibinfo {title} {Spins in Few-Electron Quantum Dots},}}
\href {https://doi.org/10.1103/RevModPhys.79.1217}
{\bibfield  {journal} {\bibinfo  {journal} {Rev. Mod. Phys.}}
	\textbf {\bibinfo {volume} {79}},\
	\bibinfo {pages} {1217} (\bibinfo {year} {2007})}
\BibitemShut {}

\bibitem [{\citenamefont {Hayashi et~al.}(2003)}]{Hayashi2003}%
\BibitemOpen
\bibfield  {author} {\bibinfo {author} {\bibfnamefont {T.}~\bibnamefont {Hayashi}},}
\bibfield  {title} {\enquote {\bibinfo {title} {Coherent Manipulation of Electronic States in a Double Quantum Dot},}}
\href {https://doi.org/10.1103/PhysRevLett.91.226804}
{\bibfield  {journal} {\bibinfo  {journal} {Phys. Rev. Lett.}}
	\textbf {\bibinfo {volume} {91}},\
	\bibinfo {pages} {226804} (\bibinfo {year} {2003})}
\BibitemShut {}

\bibitem [{\citenamefont {Fujisawa et~al.}(1998)}]{Fujisawa1998}%
\BibitemOpen
\bibfield  {author} {\bibinfo {author} {\bibfnamefont {T.}~\bibnamefont {Fujisawa}},}
\bibfield  {title} {\enquote {\bibinfo {title} {Spontaneous Emission Spectrum in Double Quantum Dot Devices},}}
\href {https://doi.org/10.1126/science.282.5390.932}
{\bibfield  {journal} {\bibinfo  {journal} {Science}}
	\textbf {\bibinfo {volume} {282}},\
	\bibinfo {pages} {932} (\bibinfo {year} {1998})}
\BibitemShut {}

\bibitem [{\citenamefont {Paladino et~al.}(2014)}]{Paladino2014}%
\BibitemOpen
\bibfield  {author} {\bibinfo {author} {\bibfnamefont {E.}~\bibnamefont {Paladino}},}
\bibfield  {title} {\enquote {\bibinfo {title} {1/f Noise: Implications for Solid-State Quantum Information},}}
\href {https://doi.org/10.1103/RevModPhys.86.361}
{\bibfield  {journal} {\bibinfo  {journal} {Rev. Mod. Phys.}}
	\textbf {\bibinfo {volume} {86}},\
	\bibinfo {pages} {361} (\bibinfo {year} {2014})}
\BibitemShut {} 

\bibitem [{\citenamefont {Petersson et~al.}(2010)}]{Petersson2010}%
\BibitemOpen
\bibfield  {author} {\bibinfo {author} {\bibfnamefont {K.~D.}~\bibnamefont {Petersson}},}
\bibfield  {title} {\enquote {\bibinfo {title} {Charge and Spin State Readout of a Double Quantum Dot Coupled to a Resonator},}}
\href {https://doi.org/10.1038/nature08864}
{\bibfield  {journal} {\bibinfo  {journal} {Nature}}
	\textbf {\bibinfo {volume} {490}},\
	\bibinfo {pages} {380} (\bibinfo {year} {2010})}
\BibitemShut {}



\bibitem [{\citenamefont {Dial et~al.}(2013)}]{Dial2013}%
\BibitemOpen
\bibfield  {author} {\bibinfo {author} {\bibfnamefont {O.~E.}~\bibnamefont {Dial}},}
\bibfield  {title} {\enquote {\bibinfo {title} {Charge Noise Spectroscopy Using Coherent Exchange Oscillations in a Singlet-Triplet Qubit},}}
\href {https://doi.org/10.1103/PhysRevLett.110.146804}
{\bibfield  {journal} {\bibinfo  {journal} {Phys. Rev. Lett.}}
	\textbf {\bibinfo {volume} {110}},\
	\bibinfo {pages} {146804} (\bibinfo {year} {2013})}
\BibitemShut {}



\bibitem [{\citenamefont {Bylander et~al.}(2011)}]{Bylander2011}%
\BibitemOpen
\bibfield  {author} {\bibinfo {author} {\bibfnamefont {J.}~\bibnamefont {Bylander}},}
\bibfield  {title} {\enquote {\bibinfo {title} {Noise Spectroscopy through Dynamical Decoupling with a Superconducting Qubit},}}
\href {https://doi.org/10.1038/nphys1994}
{\bibfield  {journal} {\bibinfo  {journal} {Nat. Phys.}}
	\textbf {\bibinfo {volume} {7}},\
	\bibinfo {pages} {565} (\bibinfo {year} {2011})}
\BibitemShut {}



\bibitem [{\citenamefont {Liu et~al.}(2011)}]{Liu2011}%
\BibitemOpen
\bibfield  {author} {\bibinfo {author} {\bibfnamefont {B.-H.}~\bibnamefont {Liu}},}
\bibfield  {title} {\enquote {\bibinfo {title} {Experimental Control of the Transition from Markovian to Non-Markovian Dynamics of Open Quantum Systems},}}
\href {https://doi.org/10.1038/nphys2085}
{\bibfield  {journal} {\bibinfo  {journal} {Nat. Phys.}}
	\textbf {\bibinfo {volume} {7}},\
	\bibinfo {pages} {931} (\bibinfo {year} {2011})}
\BibitemShut {}




\makeatother
\end{thebibliography}
\end{document}